\documentclass[aps,prd,12pt,preprint,superscriptaddress,nofootinbib,floatfix]{revtex4}

\usepackage{latexsym}
\usepackage{amsfonts}
\usepackage[latin1]{inputenc}
\usepackage[caption=false]{subfig}
\usepackage{graphicx}
\usepackage{mathrsfs}
\usepackage{amssymb}
\usepackage{amsmath}
\usepackage{verbatim}
\usepackage{multirow}
\usepackage{color}
\usepackage{epsfig}
\usepackage{amscd}
\usepackage{bm}

\def\D0{\slash\!\!\!\!\!\!\!\!\!\:D0}

\begin{document}

%\begin{flushleft}
\preprint{PSI-PR-13-09, SHEP-12-32}%\\
%\begin{flushright}
\date{\today}
%\end{flushright}
%\vspace*{-0.25cm}

\title{The 4-Dimensional Composite Higgs Model (4DCHM) \\ and the 125 GeV Higgs-like signals at the LHC}

%\vspace*{1.0truecm}

\author{D.~Barducci}\email[E-mail: ]{d.barducci@soton.ac.uk}
\affiliation{School of Physics and Astronomy, University of Southampton, Southampton SO17 1BJ, U.K.}
\author{A.~Belyaev}\email[E-mail: ]{a.belyaev@soton.ac.uk}
\affiliation{School of Physics and Astronomy, University of Southampton, Southampton SO17 1BJ, U.K.}
\affiliation{Particle Physics Department, Rutherford Appleton Laboratory, Chilton, Didcot, Oxon OX11 0QX, UK}
\author{M.~S.~Brown}\email[E-mail: ]{m.s.brown@soton.ac.uk}
\affiliation{School of Physics and Astronomy, University of Southampton, Southampton SO17 1BJ, U.K.}
\author{S.~De~Curtis}\email[E-mail: ]{decurtis@fi.infn.it}
\affiliation{INFN, Sezione di Firenze, Via G. Sansone 1, 50019 Sesto Fiorentino, Italy}
\author{S.~Moretti}\email[E-mail: ]{s.moretti@soton.ac.uk}
\affiliation{School of Physics and Astronomy, University of Southampton, Southampton SO17 1BJ, U.K.}
\affiliation{Particle Physics Department, Rutherford Appleton Laboratory, Chilton, Didcot, Oxon OX11 0QX, UK}
\author{G.~M.~Pruna}\email[E-mail: ]{giovanni-marco.pruna@psi.ch}
\affiliation{TU Dresden, Institut f\"ur Kern- und Teilchenphysik, Zellescher Weg 19, D-01069 Dresden, Germany}
\affiliation{Paul Scherrer Institute, CH-5232 Villigen PSI, Switzerland}

\begin{abstract}
\noindent
We show that the 4-Dimensional Composite Higgs Model (4DCHM) could provide an even better explanation than the Standard Model (SM) of  the current Large Hadron Collider (LHC) data pointing to the discovery of a neutral Higgs boson. The full particle spectrum of this scenario is derived without any approximation and implemented in automated computational tools to enable fast phenomenological investigation. Several benchmark sets compliant with the aforementioned data are presented and discussed.
\end{abstract}

\maketitle
%\tableofcontents

\newpage

%%%%%%%%%%%%%%%%%%%%%%%%%%%%%%%%%%%%%%%%%%%%%%%%%%%%%%%%%%

\section{Introduction}
\label{Sec:Intro}
\noindent

The discovery of a new bosonic resonance  is the first exciting outcome of the worldwide long-term effort of the particle physics community built around the Large Hadron Collider (LHC). Although the properties of the new particle have not been determined yet, after the recent updates from the ATLAS and CMS experiments \cite{:2012gk,:2012gu}\footnote{Some moderate evidence in this direction also emerged at the Tevatron \cite{Aaltonen:2012qt}.}, the general consensus is narrowing down on the common ground that a Standard Model (SM) like Higgs boson has been discovered, with a mass around 125--126 GeV. 
{We also have some hints on possible deviations from the SM Higgs expectations in the $H\to \gamma\gamma$ channel, where the number of observed events  is somewhat higher than what is predicted by the SM, whereas the $WW$, $ZZ$ rates are roughly consistent with it and the  $b\bar b$ and $\tau^+\tau^-$ ones are instead rather inconclusive, as the errors are here still rather large.} Finally, no statement yet can be made about the nature of the new object, whether fundamental or composite.

In composite Higgs models, such a state can emerge as a Pseudo-Nambu-Goldstone Boson (PNGB) from a strongly-interacting sector, which generally comprises both new bosonic and fermionic resonances \cite{Dimopoulos:1981xc,Banks:1984gj,Georgi:1984af,Georgi:1984ef,Kaplan:1983fs,Kaplan:1983sm,Dugan:1984hq}. Its production and decay dynamics are no longer controlled by just one quantity, e.g. its mass, rather additional parameters are responsible for the Higgs boson properties, which can in turn deviate from the SM ones. This is true in particular for the case of the production and decay channels exploited in the LHC search. It is then of importance to review the extent to which such an hypothesis is compatible with experimental data. 

The aforementioned new states, in order to comply with current experimental bounds, ought to be heavier than such a composite Higgs state, so that at present collider energies the apparent particle content of models comprising such a new scalar state is essentially the one of the SM itself. However, such new states can affect Higgs dynamics through mixing effects: i.e., typically, new $W'$, $Z'$ gauge bosons mix with the SM ones, $W$, $Z$, and new $b'$, $t'$-type fermions  (e.g. with charge -1/3 and 2/3 respectively) mix with the  SM ones $b,t$-type. (In fact, in the spectrum of such models also heavy fermions with exotic electromagnetic charges (e.g. 5/3 and -4/3) do exist, although they do not couple directly to the Higgs boson.)  { Due to these mixings and also to deviations
arising from the fact that the $SU(2)\times U(1)$ symmetry embedding the PNGB is non-linearly
realised (see later on), Higgs couplings to such SM states are in the end modified.}
 
A flurry of literature on the subject has therefore concentrated upon studying how the current searches for a Higgs boson are affected by the modifications of its couplings in various composite Higgs models \cite{Falkowski:2007hz,Azatov:2011qy,Azatov:2012rd,Gillioz:2012se,Azatov:2012ga}. In this connection, {one should note that} the approach that has generically been taken so far, the one of essentially accounting for the new heavy states in their infinite mass limit and studying the residual effects in the SM sector{\footnote{{With the exception of Ref.~\cite{Gillioz:2012se}, in which mass effects of the new fermionic partners have been taken into account and discussed in the case of single and double Higgs boson production via gluon-gluon fusion and of its 
di-photon decays.}}}, may not be sufficiently accurate if one notices that the heavy quarks could be not so much heavier than the presumed Higgs mass. On the one hand, both species can affect the mixing pattern of the Higgs boson, modifying its couplings to the SM particles. On the other hand,  they appear as virtual objects interacting with the Higgs boson active at the LHC, if one realises that the Higgs production channel to which the LHC is most sensitive for a mass around 125--126 GeV is gluon-gluon fusion (which can occur in such models via not only loops of $t$, $b$ quarks but also via $t'$, $b'$ ones) and that the decay channel that appears most anomalous is the photon-photon one (which can occur in such models via not only loops of $t$, $b$ quarks and $W$ bosons but also via $t'$, $b'$ and $W'$ ones). In essence, it is clear that a more rigorous approach may be needed.

It is the purpose of this paper to go beyond the approximation commonly used  in such composite Higgs models, which generally accounts for  modified Higgs couplings (to SM objects) whilst treating the heavy bosonic and fermionic states as essentially decoupled (i.e., in their infinite mass limit). We will prove that the exact results can deviate from this hypothesis by adopting a particular composite Higgs model for which we have derived exactly the spectrum of particle masses and couplings which would intervene at current and future LHC energies. 
%Yet, we will argue that such effects would
%persist in the entire class of such scenarios.

The paper is organised as follows. In the next section, \ref{Sec:Model}, we introduce the reference model adopted and describe at some length its Higgs sector. In Sect.~\ref{Sec:strategy} we extract its allowed parameter space in the light of the latest experimental findings. In Sect.~\ref{Sec:Results} we present our results. Finally, we conclude in Sect.~\ref{Sec:Conclusions}. We also have two appendices (\ref{appe:a} and \ref{appe:b}) which give detail of the loop calculations entering our analysis.

%%%%%%%%%%%%%%%%%%%%%%%%%%%%%%%%%%%%%%%%%%%%%%%%%%%%%%%%%%

\section{4DCHM: the Higgs sector}
\label{Sec:Model}
\noindent

Here  we will very briefly recap the key features of the 4DCHM introduced in \cite{DeCurtis:2011yx}, to which we refer for further details throughout this section. Our main interest will concern  the composite Higgs particle which is predicted to be  a PNGB and its couplings to both the SM particles (mainly to the $W$ and $Z$ gauge bosons plus the $t$ and $b$ quarks) and to the other new particles described by the model (the $W'$ and $Z'$ gauge bosons plus  the $t'$ and $b'$ quarks).

The 4DCHM is an effective low-energy Lagrangian approximation that represents an extremely deconstructed version of the Minimal Composite Higgs model (MCHM) of \cite{Agashe:2004rs} based on the coset $SO(5)/SO(4)$ that gives four GBs in the vector representation of $SO(4)$ one of which is the composite Higgs boson. This extreme deconstruction of the 5D theory leads to a two site schematic representation, respectively called elementary and composite sectors (considered already in  \cite{Contino:2006nn} where, however, the full gauge/Goldstone boson structure of the theory is not incorporated). Although extreme, this two site truncation represents the framework where to study in a computable  way the lowest lying resonances (both bosonic and fermionic) that are the only ones that may be accessible at the LHC. In essence, the 4DCHM represents the ideal phenomenological framework where to test the idea of a composite Higgs boson as a PNGB (see also \cite{Panico:2011pw} although with a different construction).

For a detailed analysis of the phenomenology of the gauge sector of the 4DCHM we refer to \cite{Barducci:2012kk} and \cite{Barducci:2012as}, where the Drell-Yan (DY) and di-boson processes (both with leptonic final states) have been studied, as well as to \cite{Barducci:2012sk}, where $t\bar t$ hadron-production was tackled, while here we present the main characteristics of the Higgs sector.

In the 4DCHM the physical Higgs particle acquires mass through a one-loop generated potential (the Coleman-Weinberg one). The particular choice for the fermionic sector of \cite{DeCurtis:2011yx}, one of the most economic ones in the class of composite Higgs models, gives a finite potential and from the location of the minimum one extracts the expression for the mass of the Higgs boson, $m_H$, and its VEV, $\langle h \rangle$, in terms of the parameters of the model. Further, for a natural choice of these, the Higgs mass can be consistent with the recent results of the CMS \cite{:2012gu} and ATLAS \cite{:2012gk} experiments, measuring $m_H$ around $125$ GeV. Also for this reason then we will adopt the effective description of the 4DCHM for our phenomenological analysis of Higgs processes at the LHC.

In the spirit of partial compositeness, spin 1 and spin 1/2 particles from the SM are coupled to the Higgs boson only via mixing with the composite particles while the new gauge and fermionic resonances directly interact with the Higgs field \footnote{{In the 4DCHM, regarding the fermionic sector, only the top and bottom quarks are mixed with the composite fermions. However, it is possible to extend this feature also to the other generations of quarks and to the leptons: see, e.g., \cite{Carmona:2013cq}.}}. In order to write  the interaction Lagrangian of the Higgs boson with both the gauge bosons and the fermions,  we must recall the gauge and fermionic Lagrangian of the 4DCHM, namely:
\begin{equation}
 \begin{split}
  \mathcal{L}_{gauge}&=\frac{f_1^2}{4}Tr|D_{\mu}\Omega_1|^2+\frac{f_2^2}{2}(D_{\mu}\Phi_2)(D_{\mu}\Phi_2)^T-
  \frac{1}{4}\rho_{\mu\nu}^{\tilde A}\rho^{{\tilde A}\mu\nu}-\frac{1}{4}F_{\mu\nu}^{\tilde W} F^{{\tilde W}\mu\nu} \\ 
  \mathcal{L}_{fermions}&=\mathcal{L}_{fermions}^{el}+ (\Delta_{t_L}\bar{q}^{el}_L\Omega_1\Psi_T+\Delta_{t_R}\bar{t}^{el}_R\Omega_1\Psi_{\tilde{T}}+h.c.)\\
  &+\bar{\Psi}_T(i\hat{D}^{\tilde{A}}-m_*)\Psi_T+\bar{\Psi}_{\tilde{T}}(i\hat{D}^{\tilde{A}}-m_*)\Psi_{\tilde{T}}\\
  &-(Y_T\bar{\Psi}_{T,L}\Phi_2^T\Phi_2\Psi_{\tilde{T},R}+M_{Y_T}\bar{\Psi}_{T,L}\Psi_{\tilde{T},R}+h.c.)\\
  &+(T\rightarrow B),
 \end{split}
 \label{L4DCHM}
\end{equation}
where we can isolate its main components.

\begin{itemize}
 \item The covariant derivatives are defined with respect to the elementary fields $\tilde{W}$ and the composite ones $\tilde{A}$ that denote in a general way the mediators of the $SU(2)_L\otimes U(1)_Y$ and the new $SO(5) \otimes U(1)_X$ gauge groups with gauge couplings $g_0=\{g_0,g_{0Y}\}$ and $g_*$ (for simplicity we take equal couplings for $SO(5)$ and $U(1)_X$), respectively. Their expressions are
 \begin{equation}
 \begin{split}
  &D^{\mu}\Omega_1=\partial^{\mu}\Omega_1-i g_{0}\tilde{W}\Omega_1+i g_*\Omega_1\tilde{A},\\
  &D_{\mu}\Phi_2=\partial_{\mu}\Phi_2-i g_*\tilde{A}\Phi_2.
  \end{split}
 \end{equation}
 
 \item With the particular choice of unitary gauge the link fields $\Omega_n$ are given by
\begin{equation}
 \Omega_n={\bf{1}}+i\frac{s_n}{h}\Pi+\frac{c_n-1}{h^2}\Pi^2,~~~s_n=\sin(f h/f_n^2),~~~c_n=\cos(f h/f_n^2),  ~~~h=\sqrt{h^{\hat{a}}h^{\hat{a}}},
\end{equation}
where $\Pi=\sqrt{2} h^{\hat{a}}T^{\hat{a}}$ is the GB matrix and the $T^{\hat{a}}$'s are the $SO(5)/SO(4)$ broken generators, with $\hat{a}=1,2,3,4$. 
{Therefore, the explicit expression for $\Pi$ is given by}
\begin{equation}
\Pi=\sqrt{2}h^{\hat{a}}T^{\hat{a}}=-i\left( \begin{array}{cc}
0_4 & \textbf{h}  \\
-\textbf{h}^T & 0  \\ 
\end{array}\right),\quad \textbf{h}^T=\left( h_1, h_2, h_3, h_4 \right).
\label{h4}
\end{equation}

It is also possible to relate {\bf{h}} to the usual  SM $SU(2)_L$ Higgs doublet via the following relation
\begin{equation}
H=\frac{1}{\sqrt{2}}\left( \begin{array}{c}
-i h_1 - h_2 \\
-i h_3 + h_4  \\ 
\end{array}\right).
\end{equation}

\item The $f_{i}$'s are the link coupling constants and $f$ the strong sector scale, which, in the unitary gauge, are related through
\begin{equation}
\sum_{n=1}^2\frac{1}{f_n^2}=\frac{1}{f^2}.
\end{equation}

\item The field $\Phi_2$ is a vector of $SO(5)$  that describes the spontaneous symmetry breaking of $SO(5)\otimes U(1)_X \rightarrow SO(4)\otimes U(1)_X$ and is defined as
\begin{equation}
\Phi_2=\phi_0\Omega_2^T \quad \text{where} \quad \phi_0^i=\delta^{i5}.
\end{equation}
\item The last two entries in $\mathcal{L}_{gauge}$ are the kinetic terms of  the composite and elementary gauge fields, respectively.

\item $\Psi_{T,B}$ and $\tilde{\Psi}_{T,B}$ are the fundamental representations of $SO(5)\otimes U(1)_X$ in which the new fermionic resonances are embedded while the SM third generation quarks are embedded in an incomplete representation of $SO(5)\otimes U(1)_X$ in such a way that the relation $Y=T^{3R}+X$ can reproduce the correct quantum numbers of these quarks under the $SU(2)_L \otimes U(1)_Y$ structure of the SM.

\item $\Delta_{t,b/L,R}$ are the mixing parameters between the elementary and composite sectors whereas $Y_{T,B}$ and $M_{Y_{T,B}}$ are Yukawa parameters of the composite sector and $m_*$ is the mass parameter of the new fermionic resonances.

\item {{$\mathcal{L}_{fermions}^{el}$ is the term describing the fermionic kinetic terms and the interactions of the two lightest generation of quarks and the three generations of leptons with the elementary gauge fields.}}
\end{itemize}

With these definitions and by using
\begin{equation}
 \Omega_n={\bf{1}}+\delta \Omega_n
\end{equation}
we can write the interaction of the Higgs field (which, with our choice of the vacuum alignment, is identified with $H=h_4$ in eq.~(\ref{h4})) with the gauge and fermionic fields, that are described by the following Lagrangian:
\begin{equation}
 \begin{split}
\mathcal{L}_{gauge,H}=&-\frac{f_1^2}{2}g_0 g_* Tr\left[\tilde{W}\delta\Omega_1 \tilde{A}+\tilde{W}\tilde{A}\delta\Omega_1^T+\tilde{W}\delta\Omega_1 \tilde{A} \delta\Omega_1^T   \right]\\
&+\frac{f_2^2}{2}g_*^2\Big[\phi_0^T\delta\Omega_2^T\tilde{A}\tilde{A}\phi_0+\phi_0^T\tilde{A}\tilde{A}\delta\Omega_2\phi_0
+\phi_0^T\delta\Omega_2^T\tilde{A}\tilde{A}\delta\Omega_2\phi_0\Big ],\\
\mathcal{L}_{ferm,H}=&\Delta_{t_L}\bar{q}^{el}_L\delta\Omega_1\Psi_T+\Delta_{t_R}\bar{t}^{el}_R\delta\Omega_1\Psi_{\tilde{T}}\\
&-Y_T\bar{\Psi}_{T,L}(\phi_0^T\phi_0\delta\Omega_2^T+\delta\Omega_2\phi_0\phi_0^T+\delta\Omega_2\phi_0^T\phi_0\delta\Omega_2^T)\Psi_{\tilde{T},R}\\
&+(T\rightarrow B)+h.c.
\end{split}
\label{Hcoup}
\end{equation}

In the unitary gauge, the fields $h_1,~h_2,~ h_3$  are not present in the spectrum as they are "eaten" to give the physical degrees of freedom to the SM gauge bosons, and $\Omega_{1,2}$ describe only the Higgs field. In particular by expanding $\delta \Omega_{1,2}$ up to the first order in $H$, we can extract from eq.~(\ref{Hcoup}) the interaction terms of the Higgs field with the gauge bosons  and with the fermions  However the couplings to the mass eigenstates are obtained after the diagonalisation of the quadratic part of the Lagrangian in eq.~(\ref{L4DCHM}). The analytical expressions of the Higgs couplings $g_{HV_iV_j}$ and $g_{Hf_i\bar f_j}$, which we will use in the following phenomenological analysis, are very cumbersome, so we prefer to present in App. \ref{appe:b} some plots showing the behaviour of the relevant couplings as functions of the 4DCHM parameters in the gauge sector and some scans of the Higgs couplings to fermion pairs as functions of  the mass of the lightest heavy fermions.

%%%%%%%%%%%%%%%%%%%%%%%%%%%%%%%%%%%%%%%%%%%%%%%%%%%%%%%%%%

\section{Constraints on the 4DCHM parameter space}
\label{Sec:strategy}
\noindent

In addition to the SM matter and force states (the $e^-$, $\mu^-$, $\tau^-$, $\nu_{e,\mu,\tau}$ leptons, the $u$, $d$, $c$, $s$, $t$, $b$ quarks, the $\gamma$, $Z$, $W^{\pm}$, $g$ gauge bosons), the 4DCHM presents a Higgs Boson $H$, which is a PNGB, and a large number of new particles, both in the fermionic (quark) and bosonic (gauge)  sector. We summarise the additional particle content of the 4DCHM with respect to the SM in Tab.~\ref{table:partspec}.

\begin{table}[h!]
\begin{center}
\begin{tabular}{|l|l|}
\hline
Neutral Gauge Bosons & $Z_{1,2,...,5}$\\
Charged Gauge Bosons & $W_{1,2,3}^{\pm}$\\
Charge 2/3 quarks    & $T_{1,2,...,8}$ \\
Charge $-1/3$ quarks    & $B_{1,2,...,8}$ \\
Charge 5/3 quarks    & $\tilde{T}_{1,2}$ \\
Charge $-4/3$ quarks    & $\tilde{B}_{1,2}$ \\
\hline
\end{tabular}
\end{center}
\caption{Extra particle content of the 4DCHM with respect to the SM. An increasing number in the label of a particle corresponds to a larger mass of 
the particle itself.}
\label{table:partspec}
\end{table}

In order to perform an efficient phenomenological analysis of the Higgs sector we have implemented the 4DCHM in the package LanHEP \cite{Semenov:2010qt}, with the use of the SLHA+ library \cite{Belanger:2010st}, for the purpose of deriving in an automatic way the Feynman rules of the 4DCHM in CalcHEP format \cite{Pukhov:1999gg,Belyaev:2012qa}. Furthermore, we have already summarised in Tabs.~1 and 2 of Ref.~\cite{Barducci:2012kk}  the correspondence between the model {notations} used throughout this paper and the ones presented in the CalcHEP file uploaded onto the High Energy Physics Model Data-Base (HEPMDB) \cite{Brooijmans:2012yi} at {{\tt http://hepmdb.soton.ac.uk/hepmdb:1212.0120} under the name ``4DCHM''. The model we use in this paper also includes the $H\gamma\gamma$ and $Hgg$\footnote{Here, the Next-to-Leading-Order (NLO) QCD corrections are considered, although they are irrelevant in the present analysis, see Appendix~\ref{appe:a}.} loop diagrams, which we need in order to explore the Higgs boson phenomenology of the 4DCHM. Details of the analytical evaluation of these vertices and the respective Feynman diagrams are given in App.~\ref{appe:a}. This enhanced model implementation is available from the HEPMDB website at {\tt http://hepmdb.soton.ac.uk/hepmdb:0213.0123} under the name ``4DCHM (with HAA/HGG)''.}

To constrain the parameter space of the 4DCHM we have written an independent Mathematica routine \cite{mathematica}, which considers $f$ and $g_*$ as free parameters, performs a scan over $m_*$, $\Delta_{tL}$, $\Delta_{tR}$, $Y_T$, $M_{Y_T}$, $\Delta_{bL}$, $\Delta_{bR}$, $Y_B$, $M_{Y_B}$ and is able to find allowed points with respect to the physical constraints $e$, $M_Z$, $G_F$, $m_t$, $m_b$, $m_H$. The LEP, SLC, Tevatron and LHC data are taken into account by requiring $e$, $M_Z$, $G_F$ as per Particle Data Group (PDG) listing \cite{PDG} while for the top, bottom and Higgs masses we have used $165 ~{\rm GeV} \le m_t \le 175 ~{\rm GeV}$,  $2 ~{\rm GeV} \le m_b \le 6 ~{\rm GeV}$ and  $124 ~{\rm GeV} \le m_H \le 126 ~{\rm GeV}$, the latter consistent with the recent data coming from the ATLAS \cite{:2012gk} and CMS \cite{:2012gu} experiments\footnote{Notice that the top and bottom masses obtained in composite Higgs models have to be run down from the composite scale, so that the mass intervals  adopted here for $m_t$ and $m_b$ reflect the uncertainties entering such an evolution.}. {Additionally, bounds from Electro-Weak Precision Tests (EWPTs) were enforced by requiring the mass of the extra gauge
bosons of the order 2 TeV or larger (see \cite{Barducci:2012kk} for details).} Further notice that we have compared the $W^-t\bar b$, $Zt\bar t$ and $Z b\bar b$ couplings as well to data. In particular, regarding the latter, our program also checks that the left- and right-handed couplings of the $Z$ boson to the bottom (anti)quark are separately consistent with results of LEP and SLC \cite{Z-Pole}. For additional details on the scanning procedure, we refer the reader to \cite{Barducci:2012kk}. In a more general framework the partial compositeness should be considered for all the quark generations and therefore it should be mandatory to analyse in detail the constraints from flavour physics. However, this is beyond the scope of our analysis, hence we will not further discuss such constraints.

As mentioned, in the 4DCHM description, additional fermions  belong to its spectrum, to which we collectively refer as $t'$ and $b'$ for the ones with SM-like charges.  As these states are heavy quarks, they can potentially be produced in hadron-hadron collisions. The most stringent limits on their mass come presently from the LHC.  In order to take into account the latter, an analysis of the compatibility of the 4DCHM with LHC direct measurements has been performed. The pair production cross section $\sigma(pp\rightarrow t' \bar t'/b'\bar b')$ has been calculated according to the code described in \cite{Cacciari:2011hy}, which is essentially the one generally used to emulate $t\bar t$ production. Clearly, in the 4DCHM, such mass limits would apply to the lightest $t'$ and $b'$ states, i.e., $T_1$ and $B_1$ in Tab.~\ref{table:partspec}.

\begin{figure}[!t]
%\centering
\subfloat[]{
\includegraphics[width=0.45\linewidth]{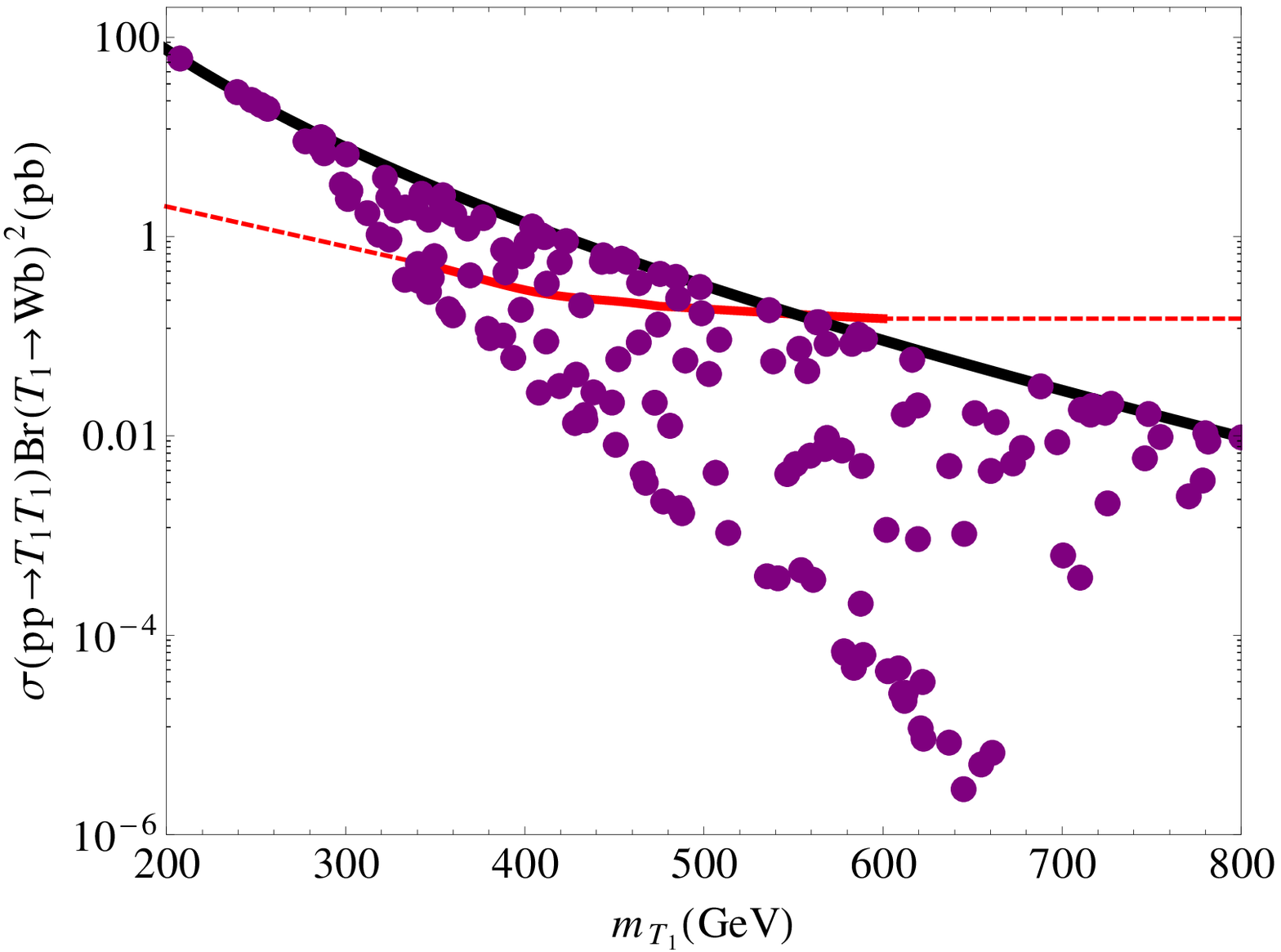}}
\subfloat[]{
\includegraphics[width=0.45\linewidth]{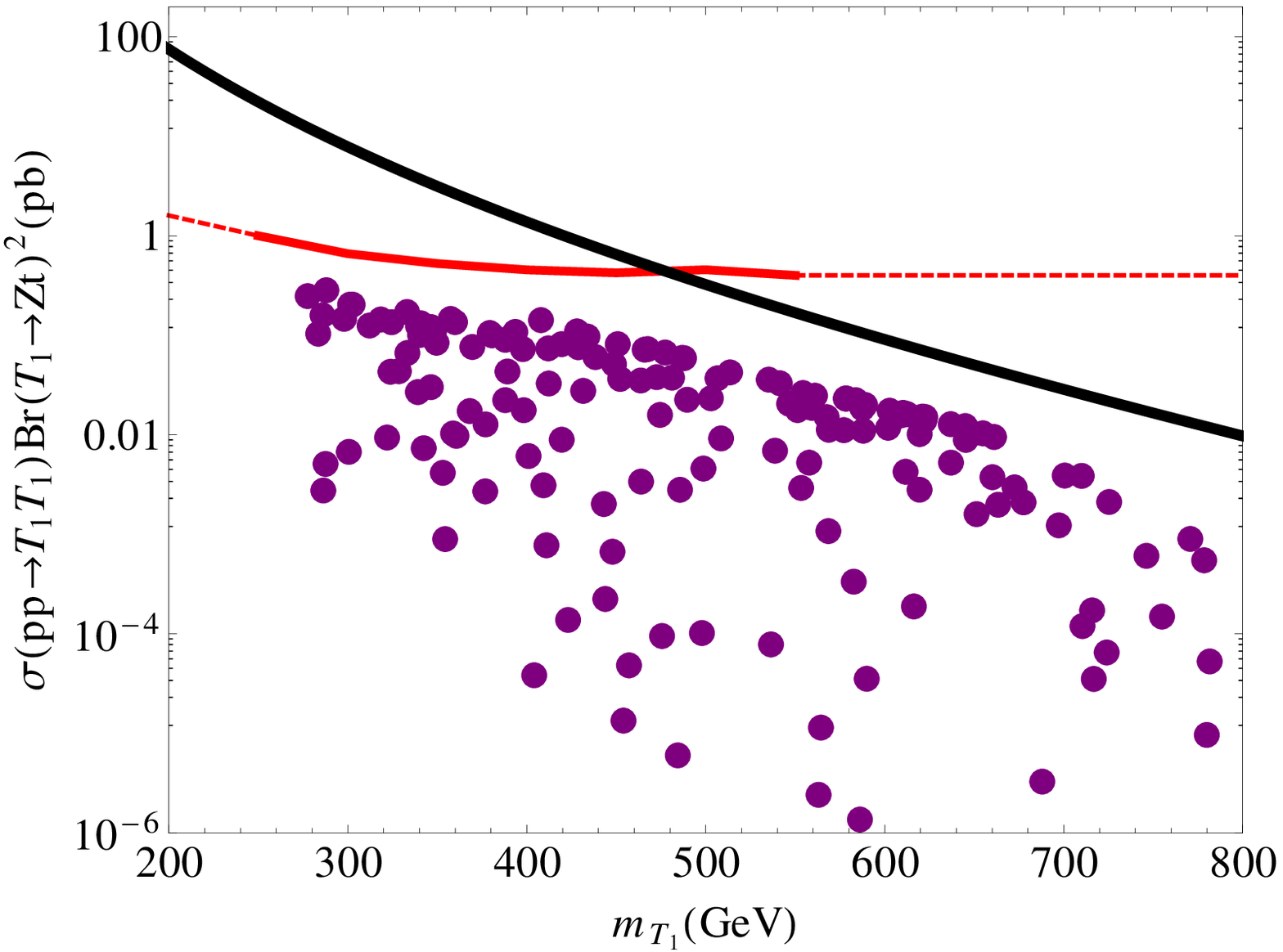}}
\caption{Limits on the mass of $m_{T_1}$ with respect to a search for $t'$s at the CMS detector with 5 fb$^{-1}$ of integrated luminosity, that assumes that the $t'$ decays 100\% in $W^+b$ (a) and 1.14  fb$^{-1}$ of integrated luminosity, that assumes that the $t'$ decays 100\% in $Zt$ (b). The black line is the cross section assuming 100\% of decay in $W^+b/Zt$, red line is the 95\% CL observed limit and the purple circles are the 4DCHM points obtained for $f=1$  TeV and $g_*=2$. The dotted-red line corresponds to our extrapolation of the experimental result (see the text). \label{fig:limit-tp}}
\end{figure}

\begin{figure}[!t]
\centering
\subfloat[]{
\includegraphics[width=0.45\linewidth]{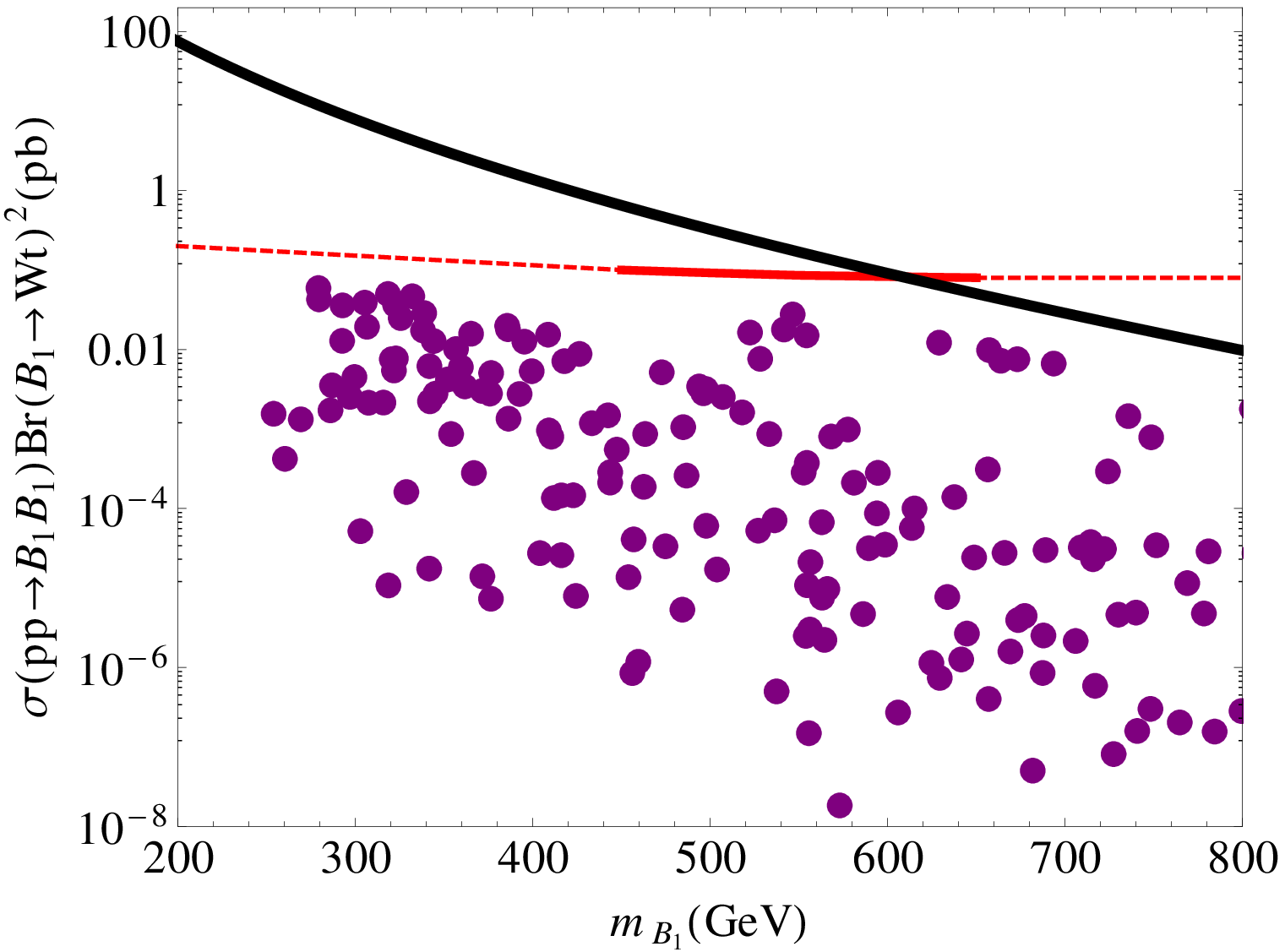}}
\subfloat[]{
\includegraphics[width=0.45\linewidth]{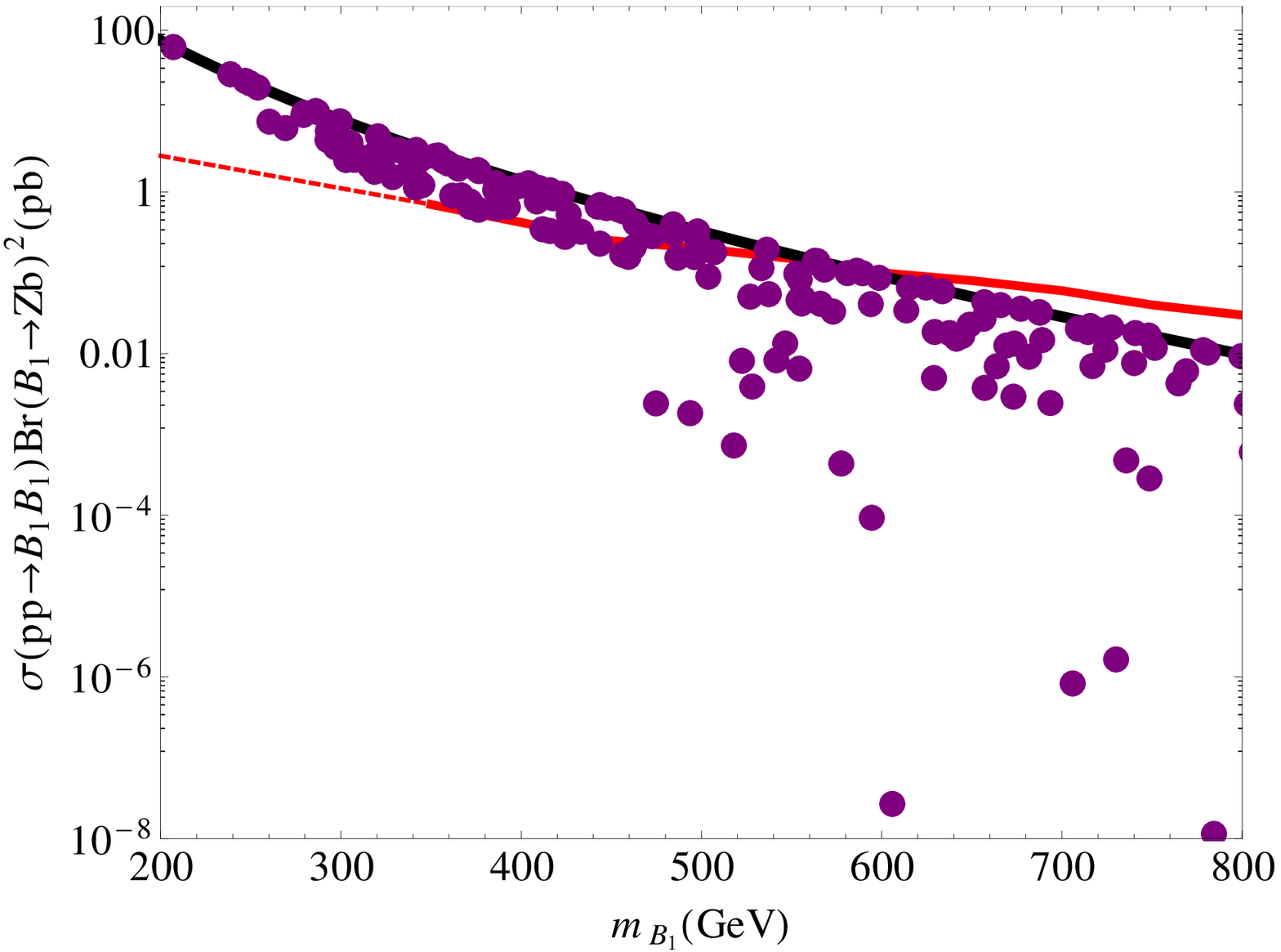}}
\caption{Limits on the mass of $m_{B_1}$ with respect to a search for $b'$'s at the CMS detector with 4.9  fb$^{-1}$ of integrated luminosity, that assumes that the $b'$ decays 100\% in $Wt$ (a) and 4.9  fb$^{-1}$ of integrated luminosity, that assumes that the b' decays 100\% in $Zb$ (b). The black line is the cross section assuming 100\% of decay in $W^+t/Zb$, red line is the 95\% CL observed limit and the purple circles are the 4DCHM points obtained for $f=1$  TeV and $g_*=2$. The dotted-red line corresponds to our extrapolation of the experimental result (see the text). \label{fig:limit-bp}}
\end{figure}

The limits on $t'$s are based on \cite{CMS:2012ab}, where a search for pair production of $t'$s is performed in CMS with 5 fb$^{-1}$ of  luminosity, where the $t'$s are assumed to decay 100\% into $W^+b$, and on \cite{Chatrchyan:2011ay}, where a search for pair production of $t'$s is performed at CMS with 1.14 fb$^{-1}$ of  luminosity, where the $t'$'s are assumed to decay 100\% into $Zt$. The limits on $b'$s are based on \cite{Chatrchyan:2012yea}, where a search for pair production of $b'$s is performed at CMS  with 4.9 fb$^{-1}$ of luminosity with the $b'$'s that are assumed to decay 100\% into $W^-t$, and on \cite{CMS-PAS-EXO-11-066}, where a search for pair production of $b'$s is performed at CMS with 4.9  fb$^{-1}$ of luminosity with the $b'$'s that are assumed to decay 100\% into $Zb$.  Finally, notice that data considered here come from the 7 TeV run of the LHC. Results for $T_1$ and $B_1$ are shown in Figs.~\ref{fig:limit-tp} and \ref{fig:limit-bp}, respectively\footnote{{More recent results from CMS are given in \cite{Chatrchyan:2012vu} and \cite{Chatrchyan:2012af}, however, they do not change our forthcoming conclusions.}}.


{We would now like to make a few remarks on the extrapolation of the experimental limits on  $m_{T_1}$ and $m_{B_1}$ used in the figures. First of all,  to be on the conservative side, we have used a linear extrapolation on the lower end of the experimental limit. This extrapolated constraint is expected to be better than the experimental one, therefore it conservatively excludes a  bigger part of the 4DCHM parameter space. The same argument and approach is valid for our extrapolation on the upper end of the experimental limit, as we have simply extended the latter horizontally. Furthermore, we would like to encourage the LHC experimental groups to explore and provide limits for a wider range of the heavy quark masses, which would eventually be useful for testing a wider expanse of the parameter space of the 4DCHM (and in fact other models as well).}
 
Before proceeding to extract mass limits from these plots, it is important to note that in the 4DCHM the bounds on the masses of the lightest top and bottom heavy partners ($T_1$ and $B_1$, respectively) are strictly correlated, the two being function of a common parameter of the model, $m_*$, so a higher mass bound on, say, $B_1$ reflects also on $T_1$. {This correlation is not taken into account in our present analysis, however, we expect it to be inconsequential for our purposes, owing to the fact that discrepancies between exclusion limits obtained from correlated and un-correlated searches would amount to no more than the actual mass difference between $T_1$ and $B_1$, generally of a few tens of GeV,
 thus being of modest relevance for the loop effects we are concerned with.} More importantly, herein, a rescaling of the $t'$ and $b'$ production cross section to account for the non-100\% Branching Ratios (BRs) of the $t'$ and $b'$ states into SM-like decay channels, owing to the new ones specific to the 4DCHM, has been taken into account {{so that this procedure may allow for the presence of $T_1$ and/or $B_1$ states with mass below the experimental exclusion limit, which indeed assumes 100\% BR in one SM-like  channel only.}} {A typical lower limit on both $m_{T_1}$ and $m_{B_1}$ of 400 GeV or so is obtained for the case, e.g., $f=1$ TeV and $g_*=2$, which will be used in the remainder of the analysis for illustrative purpose. Notice that
we obtain a lower limit on both $m_{T_1}$ and $m_{B_1}$ by always combining all the results in Figs.~\ref{fig:limit-tp}--\ref{fig:limit-bp}, namely, we require that
each point in the parameter space lies below the four experimental limit lines simultaneously. Therefore}, whenever we will present results for other benchmark values of these parameters, corresponding limits on $m_{T_1}$ and $m_{B_1}$ will be enforced.

{{Furthermore, one ought to notice that, beside the heavy fermions with ordinary charges, i.e., the $t^{\prime}$ and $b^{\prime}$
states, the composite fermionic spectrum presents also states with exotic charge, as mentioned in the model description. Although these states do not couple directly to the Higgs boson, and so they are inert for the purpose of studying 
loop-induced Higgs production and decay modes, 
it is important to set bounds on their masses since, in certain region of the parameter space, they can be almost degenerate with the lightest $t'$ or  $b'$.  Since the fermionic spectrum is determined by the  parameters we listed in Sect. II, it is clear that a bound on the $\tilde T_1$ (the lightest fermion with  charge $5/3$) mass, reflects also on $m_{T_1}$ and $m_{B_1}$.
Regarding  the $\tilde T_1$, since in the 4DCHM this particle decays almost 100\% of the times into $W^+t$, it is possible to apply directly the bound of $650\;{\rm GeV}$ given by \cite{ATLAS:2012hpa}.
Nevertheless, there are regions of the fermion parameter space where the $\tilde T_1$ is not the lightest heavy fermion. This means that values of $m_{T_1}$ and $m_{B_1}$ around 400 GeV still survive the direct search limits.
 {{{In contrast, we have to say that, at the moment, no limits for the charge $-4/3$ fermions are given by the ATLAS and CMS collaborations. They will of course further cut on the low mass values for $T_1$ and $B_1$}}}}. 

The additional gauge bosons of the 4DCHM will be collectively denoted as $W'$ and $Z'$ and  are taken with masses (and couplings) compliant with current experimental limits from both EW precision measurements and their direct searches, as described in \cite{Barducci:2012kk}. Contrary to the case of the additional fermions though, the impact of such additional gauge states onto the Higgs sector is less relevant, as we shall comment upon later on.

%%%%%%%%%%%%%%%%%%%%%%%%%%%%%%%%%%%%%%%%%%%%%%%%%%%%%%%%%%

\section{Results}
\label{Sec:Results}
\noindent

In the previous section we have described the procedure to identify the allowed  parameter space of our 4DCHM. After this initial selection, we need to compare the yield of the surviving points with the LHC data.

A useful procedure to adopt in order to do so is to define the so-called $R$ (in other works sometimes called $\mu$) parameters, i.e., the observed signal (in terms of counted events) in a specific channel divided by the SM expectation:
%%%%%%%%%%%%%%%%%%%%%%%%
%{\bf{Starting of our editing}}
%%%%%%%%%%%%%%%%%%%%%%%%
\begin{eqnarray}\label{R}
R_{YY}=\frac{\sigma(pp\to HX)|_{\rm 4DCHM}\times {\rm BR}(H\to YY)|_{\rm 4DCHM}}{\sigma(pp\to HX)|_{\rm SM}\times {\rm BR}(H\to YY)|_{\rm SM}},
\end{eqnarray}
where $YY$ refers here to any possible Higgs decay channel and in our study we consider $YY$ =
$\gamma\gamma$, $b\bar b$,  $WW$ and $ZZ$. The particles (if any) produced in association with the Higgs boson are here denoted by $X$\footnote{In reality, one should notice that eq.~(\ref{R}) is the limiting case in which sensitivity to the $YY$ decay channel is through only one of the production processes \cite{Spira:1995rr,Kunszt:1996yp,LHCHiggsCrossSectionWorkingGroup:2011ti,Heinemeyer:2013tqa}. One should more accurately sum over all of the latter. However, given present data samples and for our purposes, such an approximation is sufficiently accurate.}. For the latest experimental results on such quantities, wherein the label 4DCHM is meant to signify actual experimental data, see Tab.~\ref{tab:R}. 

The relevant hadro-production processes at partonic level are
\begin{equation}\label{gg}
gg\to H
\quad\quad ({\rm gluon-gluon~fusion}),
\end{equation}
\begin{equation}\label{VBF}
q\bar q\to q\bar q H
\quad\quad ({\rm vector~boson~fusion}),
\end{equation}
\begin{equation}\label{HS}
q\bar q(') \to V H
\quad\quad ({\rm Higgs-strahlung}),
\end{equation}
where $V=W,Z$. Notice that 
\begin{equation}\label{QQH}
gg,q\bar q\to Q\bar Q H
\quad\quad ({\rm associate~quark~production})
\end{equation}
(for $Q=b,t$) is generally negligible at the energies so far attained by the LHC, of 7 and 8 TeV.

\begin{table}[!t]
\begin{center}
\begin{tabular}{|l|l|l|}
	\hline
	 & ~~ATLAS & ~~~~CMS\\
	\hline
	$R_{\gamma\gamma}$ & $\phantom{-} 1.8 \pm 0.4\phantom{-}$ & $\phantom{-}1.564_{-0.419}^{+0.460}\phantom{-}$ \\
	$R_{ZZ}$           & $\phantom{-} 1.0 \pm 0.4\phantom{-}$ & $\phantom{-}0.807_{-0.280}^{+0.349}\phantom{-}$ \\
	$R_{WW}$           & $\phantom{-} 1.5 \pm 0.6\phantom{-}$ & $\phantom{-}0.699_{-0.232}^{+0.245}\phantom{-}$ \\
	$R_{bb}$		 & $- 0.4 \pm 1.0\phantom{-}$ & $\phantom{-}1.075_{-0.566}^{+0.593}\phantom{-}$ \\
	\hline
\end{tabular}
\end{center}
\caption{Summary of the LHC measurements of some  $R$ parameters  from the latest ATLAS \cite{ATLAS-CONF-2012-170} and CMS~\cite{CMS-PAS-HIG-12-045} data. (The CMS paper only presents the data in a box-plot, though the numerical values can be found in \cite{CMStwiki}.) \label{tab:R}}
%\end{ruledtabular}
\end{table}

For the purpose of our analysis, it is convenient to re-write eq.~(\ref{R}) as follows
\begin{eqnarray}\label{newR}
R_{YY}^{Y'Y'}=\frac{\Gamma(H\to Y'Y')|_{\rm 4DCHM}\times \Gamma(H\to YY)|_{\rm 4DCHM}}
                                 {\Gamma(H\to Y'Y')|_{\rm SM       }\times \Gamma(H\to YY)|_{\rm SM       }}
                         \frac{\Gamma_{\rm tot}(H)|_{\rm SM       }}
                                {\Gamma_{\rm tot}(H)|_{\rm 4DCHM}},
\end{eqnarray}
where $Y'Y'$ {denote incoming particles participating the Higgs boson production, e.g., $gg$ for process (\ref{gg}) and $VV$ for processes (\ref{VBF}) and (\ref{HS}), while $YY$ indicate particles into which the Higgs boson decays}\footnote{Notice that the coupling between the Higgs boson and $W$ or $Z$ intervening in the two production channels (\ref{VBF})--(\ref{HS}) is the same. However, also notice that, in the 4DCHM, the two couplings $HWW$ and $HZZ$ do not rescale in the same way with respect to the SM ones, in particular for the parameter space investigated here, though the differences will be shown to be small. Hereafter, we will adopt the generic label $V$ to signify either a $W$ or a $Z$. {However, we remark that in the numerical evaluation we distinguish between the Higgs couplings to $W$ and $Z$.}}.

For $YY=\gamma\gamma$, $WW$, $ZZ$ we take the dominant production process to be gluon-gluon fusion (i.e., $Y'Y'=gg$) while for $YY=b\bar{b}$ we assume that the Higgs-strahlung dominates (i.e., $Y'Y'=VV$, with the appropriate superposition of $WW$ and $ZZ$).
%%%%%%%%%%%%%%%%%%%
% remember to erase indentation
%{\bf{End of our editing}}
%%%%%%%%%%%%%%%%%%%
In other words, we trade a cross section for a width (so-to-say) and this is possible, as we will be carrying out our analysis at lowest order without the presence of radiative corrections due to either Quantum Chromo-Dynamics (QCD) or EW interactions. In fact, following Ref.~\cite{LHC-HXSWG:2012nn}, we can cast eq.~(\ref{newR}) also in the following form:
\begin{eqnarray}\label{R_vs_kappas}
R_{YY}^{Y'Y'}=\frac{\kappa^2_{Y'} \kappa^2_Y}{\kappa^2_H},
\end{eqnarray}
wherein (recall that $VV=WW$ or $ZZ$ and notice that {$Y=b/\tau/g/\gamma/V$ and $Y'=g/V$})
\begin{eqnarray}\label{kappas}
\kappa^2_{b/\tau/g/\gamma/V}=\frac{\Gamma(H\to b\bar b/\tau^+\tau^-/gg/\gamma\gamma/VV)|_{\rm 4DCHM}}
                                                                  {\Gamma(H\to b\bar b/\tau^+\tau^-/ gg/\gamma\gamma/VV)|_{\rm SM       }}
\end{eqnarray}
and
\begin{eqnarray}\label{kH}
\kappa^2_H=\frac{\Gamma_{\rm tot}(H)|_{\rm 4DCHM}}
                              {\Gamma_{\rm tot}(H)|_{\rm SM       }}.
\end{eqnarray}

The LHC experiments then perform fits to the $\kappa_i$ coefficients in order to test generic BSM assumptions (for which one or more of the the $\kappa_i$s can differ from 1). However, in most cases they fix $\kappa_H^2=1$,  assuming that the Higgs width does not change, see \cite{ATLAS-CONF-2012-170,CMS-PAS-HIG-12-045}\footnote{In fact, the only departure from this  they allow for is to take $\kappa_H^2>1$, corresponding to a value of the Higgs total width in the BSM hypothesis larger than in the SM case, thereby accounting for, e.g., invisible Higgs decays that are not captured by standard searches for the aforementioned SM-like channels.}. This is a rather restrictive condition, as most BSM models allow for values of $\kappa_H^2\ne1$, as the Higgs boson under consideration can mix, in such BSM scenarios, directly with other Higgs boson states or, else, the particles to which it couples can in turn mix. {This latter mixing (e.g., the one between $b$ quarks and/or $W,Z$ bosons, whose final states make up the predominant component of the Higgs total width for Higgs masses around 125 GeV in most such BSM scenarios, including the 4DCHM, and their heavy counterparts) affects the corresponding Higgs couplings. Such effects, whichever their nature, would induce the condition $\kappa_H<1$, as it is the case in the 4DCHM (as illustrated in App.~\ref{appe:b}). In fact, we will show later on that  many of the 4DCHM effects enter through such a modification of the Higgs total width.}

In order to illustrate the 4DCHM phenomenology, we adopt as reference point the combination $f=1$ TeV and $g_*=2$. However, the salient features extracted for this case may equally be referred to the other aforementioned benchmark points, i.e., those defined in Ref.~\cite{Barducci:2012kk}. Since the errors in Tab.~\ref{tab:R} on $b\bar b$ are very large, Fig.~\ref{fig:R} shows the correlation between the  event rate ratios of eq.~(\ref{newR}) only for the $\gamma\gamma$ and $VV$ channels\footnote{However, the $b\bar b$ case will be taken into consideration later on when making fits to data. In contrast, the $\tau^+\tau^-$ case, being even worse in the above respect, is ignored throughout.}. Furthermore, as intimated, since most of the sensitivity of the $\gamma\gamma$ and $VV$ data is to the gluon-gluon fusion production mode, which is in fact the dominant one in the 4DCHM for the parameter space tested here, like for the SM, we will neglect the effects of all others in our predictions (so that we can conveniently drop the superscripts $gg$ and $VV$ for the time being). From the plot in Fig.~\ref{fig:R} it is clear that   there is a noticeable tendency of the model to prefer $R_{\gamma\gamma}$ and $R_{VV}$ values smaller than 1  (the majority of points satisfy this condition, $R_{\gamma\gamma}$ being somewhat larger than $R_{VV}$), {with $R_{ZZ}$ showing a slightly stronger propensity than  $R_{WW}$ in this direction.} The two plotted quantities also appear to be strongly correlated, thereby hinting at a possible common origin for the 4DCHM event rate behaviour relative to the SM predictions. Moreover, it is also worth mentioning here that the rates for $R_{VV}$ in both the 4DCHM and the SM are computed for the gauge boson decay patterns  which ATLAS and CMS used in reporting the results in Tab.~\ref{tab:R}. These signatures include electrons and muons in all possible combinations entering generic `two-lepton plus missing transverse energy' and `four-lepton' signatures emerging from $WW$ and $ZZ$ pairs, respectively\footnote{This clarification is of relevance for the case of the 4DCHM, in which the $W$ and $Z$ decay rates change relatively to the SM, unlike the case of other popular BSM models. { In the 4DCHM, such modifications are currently within experimental errors}.}. {In other words, in our treatment, the decays into lepton pairs are included in the rates presented.} In addition, we have checked that the contribution of $W'$ and $Z'$ states, two of the former and three of the latter, to the yield of these final states, in both mixed 4DCHM/SM {(due to one SM and one extra virtual gauge bosons)} and pure 4DCHM channels {(due to two virtual extra gauge bosons)}, is negligible, owing to their large masses as compared to the SM gauge states $W$ and $Z$, despite their large couplings (see App.~\ref{appe:b}). Finally, notice that we have restricted ourselves here to the sample of 4DCHM points that comply with the limits on $t'$ and $b'$ states  from direct searches obtained in the previous section.

It is now useful to unfold the results in Fig.~\ref{fig:R} in terms of the three $\kappa_i^2$ entering eq.~(\ref{R_vs_kappas}), as each of these can be an independent source of variation in the 4DCHM with respect to the SM. In particular, we map such results in terms of the masses of the lightest $t'$ and $b'$ quarks, i.e., $T_1$ and $B_1$, as these are the 4DCHM quantities to which the event rate ratios are most sensitive. 

\begin{figure}[!t]
\centering
\includegraphics[width=0.40\linewidth]{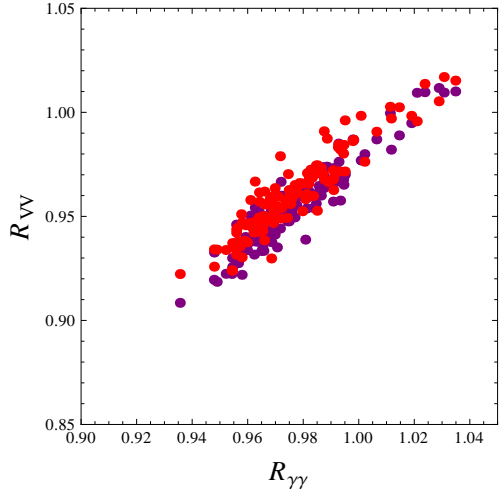}
\caption{Correlation between $R_{\gamma\gamma}$ and $R_{VV}$, with $VV=WW$ (red) and $ZZ$ (purple), from eq.~(\ref{newR}) in the 4DCHM for the benchmark point $f=1$ TeV and $g_*=2$. All points generated here are compliant with direct searches for $t'$s, $b'$s and exotic states with charge 5/3.}
\label{fig:R}
\end{figure}

We start with $\kappa_H^2$. This is shown in Fig.~\ref{fig:kH}. Herein, we keep all generated 4DCHM points, including those failing the constraints from direct searches for $t'$, $b'$ states {{or the exotic 5/3 charged fermion}}.  
This is done for the purpose of illustrating the aforementioned sensitivity of the 4DCHM predictions upon the heavy top and bottom masses. In fact, should have smaller $m_{T_1}$ and $m_{B_1}$ been allowed, effects onto the ratio of total width would have been extremely large, up to $-30\%$ or so. However, even with the aforementioned limits enforced, the 4DCHM effects induced by $t'$ and $b'$ states onto the SM remain substantial, of order $-15\%$ to $-20\%$. Hence, bearing in mind that the contribution of the $H\to gg$, $\gamma\gamma$ and $Z\gamma$ partial widths (those where such $t'$ and $b'$ states enter at lowest order) to the total one are negligible, one has to conclude that these corrections are induced by mixing effects. Furthermore, as $\Gamma_{\rm tot}(H)|_{\rm 4DCHM} \approx\Gamma(H\to b\bar b)|_{\rm 4DCHM}$ (just like in the SM), it is also clear that these are mainly due to $b'$-$b$ mixing affecting the $Hb\bar b$ coupling {(this will be shown later on, in Fig.~\ref{Hqq_b}, for a particular benchmark)}. Therefore, the result that such 4DCHM effects are negative {is not} surprising. Overall, the reduction of the total Higgs width in the 4DCHM with respect to the SM  {induces}  an increase of all $R$ values in eq.~(\ref{newR}) except, of course, $R_{b\bar b}$.

\begin{figure}[!t]
\centering
\subfloat[]{
\includegraphics[width=0.45\linewidth]{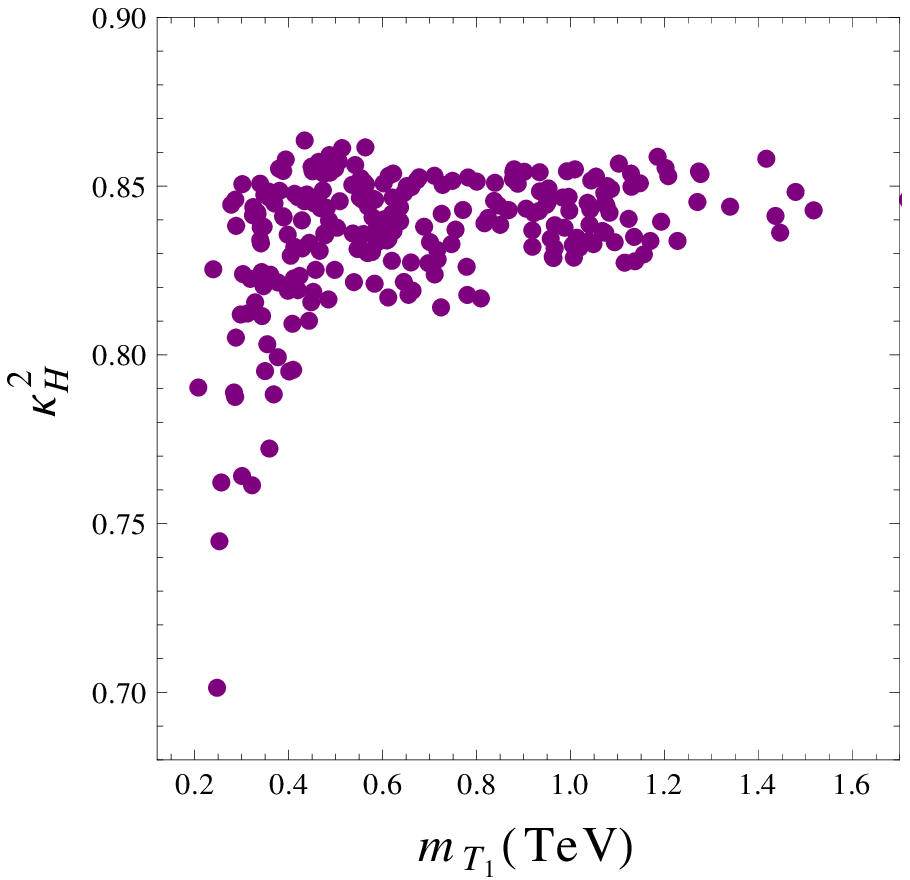}}
\subfloat[]{
\includegraphics[width=0.45\linewidth]{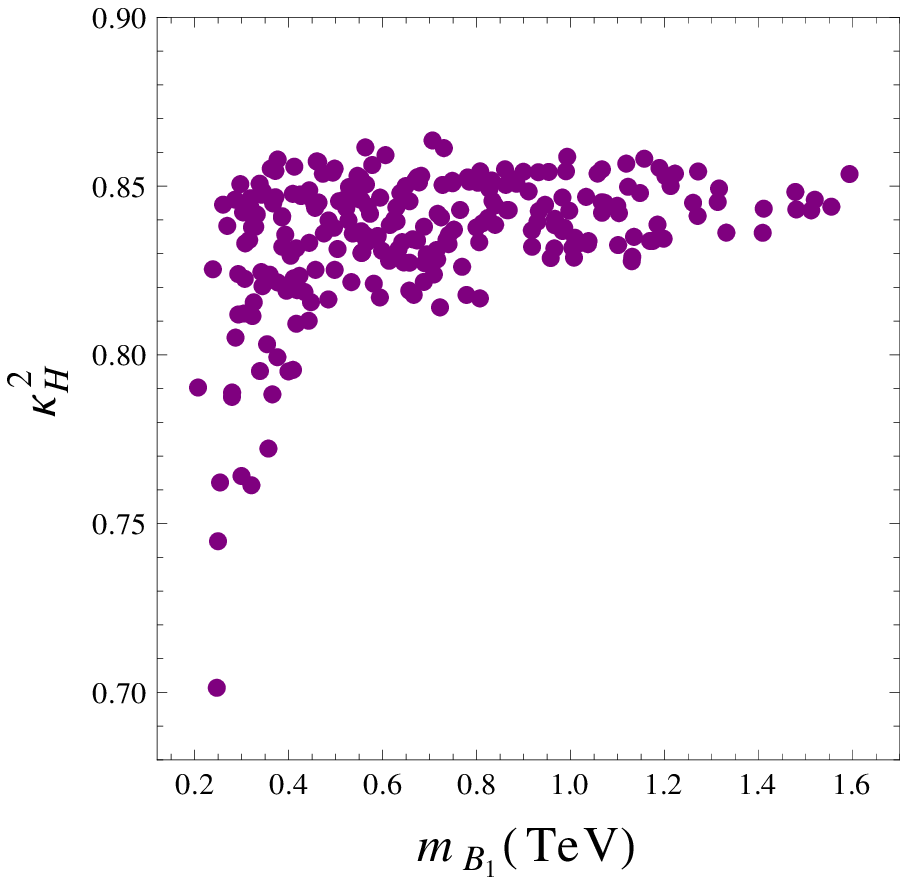}\label{fig:kH_b}}
\caption{The distributions of $\kappa_H$ values entering eq.~(\ref{R_vs_kappas}) as a function of (a) $m_{T_1}$ and (b) $m_{B_1}$ in the 4DCHM for the benchmark point $f=1$ TeV and $g_*=2$. 
%Notice that the mass regions to the left of the vertical dashed-red lines are excluded by direct searches for $t'$ and $b'$ %states. 
\label{fig:kH}}
\end{figure}

As we are interested in probing the 4DCHM hypothesis as an explanation of the LHC data used for the Higgs search and since the largest anomaly with respect to the SM is seen in the di-photon channel (recall Tab.~\ref{tab:R}), we next study $\kappa_g^2$ and $\kappa_\gamma^2$, also entering eq.~(\ref{R_vs_kappas})\footnote{The case of $\kappa_V^2$ is relevant too, as also the $HVV$ couplings in the 4DCHM change from their SM values (and, as mentioned, differently for $WW$ and $ZZ$). However, here, the dynamics occur at tree-level, so the effects are trivial, as they can be easily accounted for by replacing the $HVV$ couplings of the SM with those of the 4DCHM. Needless to say, also in this case the differences between SM and 4DCHM are negative and due to mixing, which is non-negligible, despite the fact that the masses of the heavy vector bosons, of order 1.5 TeV and more, are much larger than those of the standard gauge ones, also relatively to the case of the new fermions with respect to  the ordinary ones.}. We show this two quantities in Figs.~\ref{fig:kg}--\ref{fig:kA}, respectively. In both cases, we see a reduction of the partial widths in the 4DCHM relative to the SM. Again, we trace this to mixing effects, this time between $t$ and $t'$ states. Both at production and decay level, in fact, $t$ contributions are larger than the $b$ ones, both in the 4DCHM and SM. {Again, they induce negative corrections, typically $-10\%$ for $\kappa_{g}$ and $-6\%$ for $\kappa_{\gamma}$.} Furthermore, that the former are larger than the latter is due to the fact that the $t$ loop is the leading one in the production graph whereas it is subleading (smaller than the $W$ contribution) in the decay diagram. Incidentally, unlike the case of the total width, for these two partial widths, if lighter $T_1$ and $B_1$ masses were allowed (recall that they are strongly correlated), genuine 4DCHM effects onto the SM would have been different in the two channels, typically inducing larger(smaller) rates at production(decay) level. In this  dynamics we recognise the effects of the $t'$ and $b'$ loops in the two triangle amplitudes (as opposed to those induced by mixing in the couplings). In fact, the lighter the $t'$ and $b'$ masses, the bigger their loop contributions\footnote{Also recall that the $Ht'\bar t'$ and $Hb'\bar b'$ couplings are not of Yukawa type, that is, they do not scale linearly with the $t'$ and $b'$ masses.}. {So that, at both production and decay level, the net effect from $t'$ and $b'$ loops turn out to have the same sign as the $t$ one (recall that the $b$ ones are negligible in both models) for the case of a light $T_1$ or $B_1$ (below 500 GeV). Hence, in production they interfere constructively with the leading $t$ contribution, which in turn means that they interfere destructively in decay with the leading $W$ contribution (which has a sign opposite to the $t$ term). In case of a heavier $T_1$ and $B_1$, the sign of the overall contribution of $t'$ and $b'$ quarks can vary with respect to the top quark one but the combined contribution of all heavy quarks  is quite small. In fact, we have verified that the asymptotic values, i.e., for large $m_{T_1}$ and $m_{B_1}$, attained by $\kappa_g^2$ and $\kappa_\gamma^2$ in Figs.~\ref{fig:kg} and \ref{fig:kA}, respectively, coincide with those obtained in the aforementioned literature by adopting the described decoupling approximation of the heavy fermionic states \cite{Falkowski:2007hz,Azatov:2011qy}. Conversely, it should be noted that the asymptotic results can differ significantly from those obtained for small $T_1$ and $B_1$ masses, particularly for $\kappa_g^2$, up to $7\%$ or so (around 
400 GeV)\footnote{ Recall that any uncertainty in the computation of production and/or decay rates cancels in the $\kappa$ coefficients.}}. For smaller masses, the effect would be even more evident. We will dwell further on this in App.~\ref{appe:b}.}

\begin{figure}[!t]
\centering
\subfloat[]{
\includegraphics[width=0.45\linewidth]{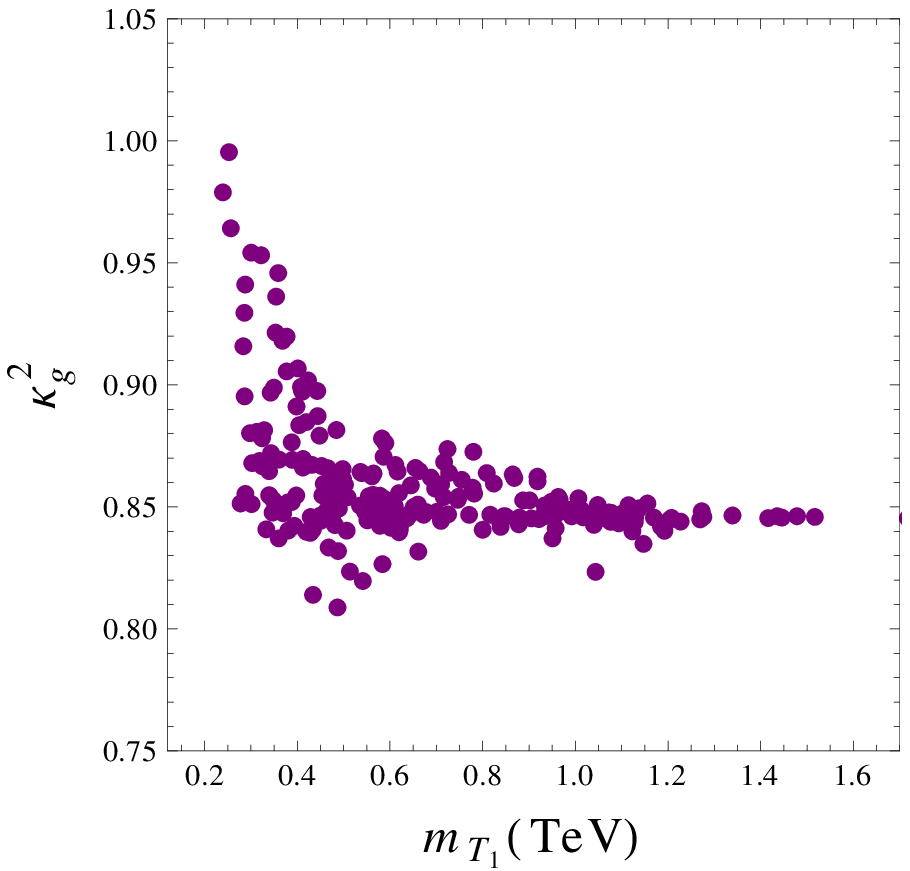}}
\subfloat[]{
\includegraphics[width=0.45\linewidth]{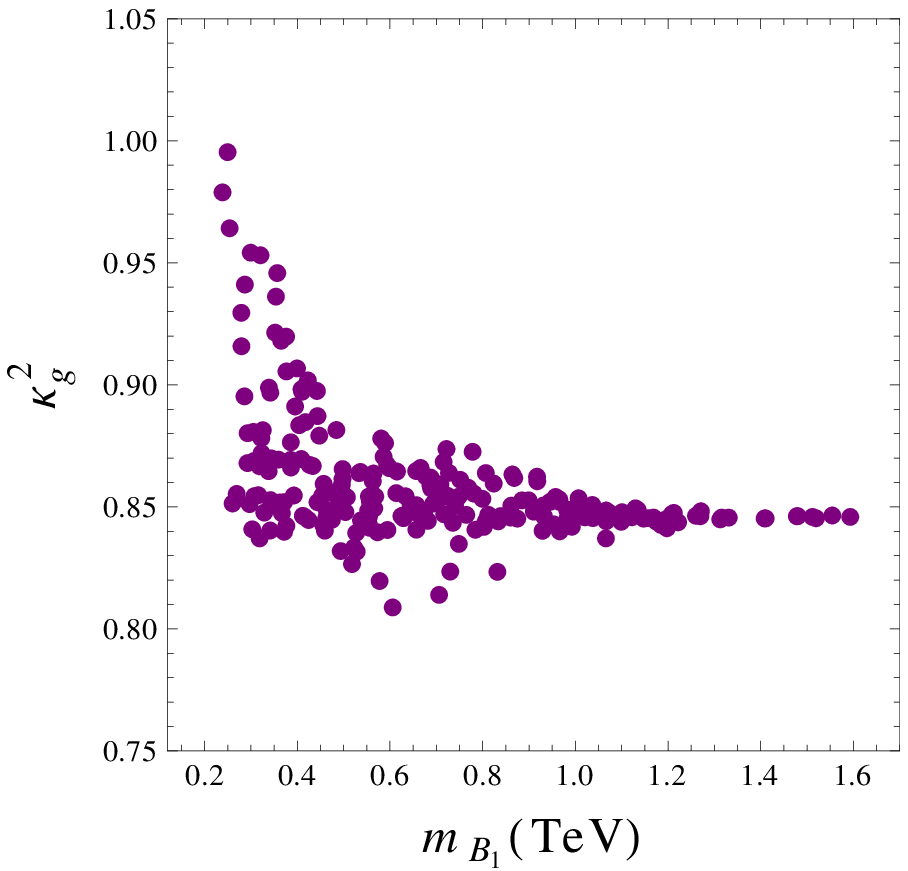}}
\caption{The distributions of $\kappa_g$ values entering eq. (\ref{R_vs_kappas}) as a function of (a) $m_{T_1}$ and (b) $m_{B_1}$ in the 4DCHM for the benchmark point $f=1$ TeV and $g_*=2$. 
%Notice that the mass regions to the left of the vertical dashed-red lines are excluded by direct searches for $t'$ and $b'$ %states. 
\label{fig:kg}}
\end{figure}

\begin{figure}[!t]
\centering
\subfloat[]{
\includegraphics[width=0.45\linewidth]{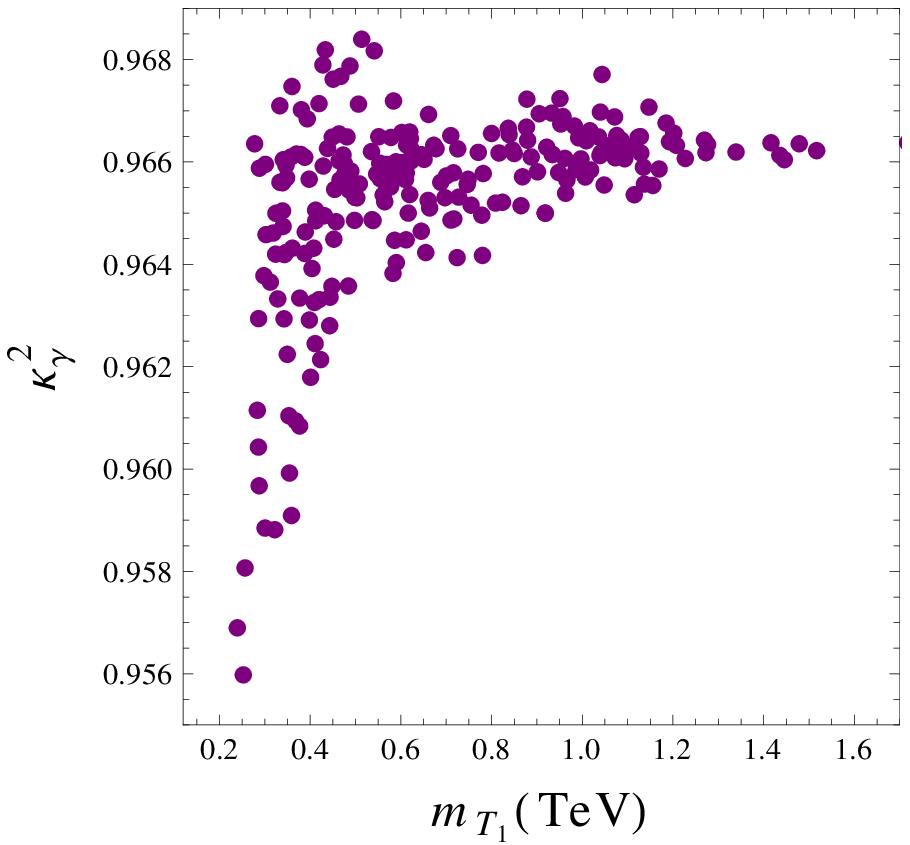}}
\subfloat[]{
\includegraphics[width=0.45\linewidth]{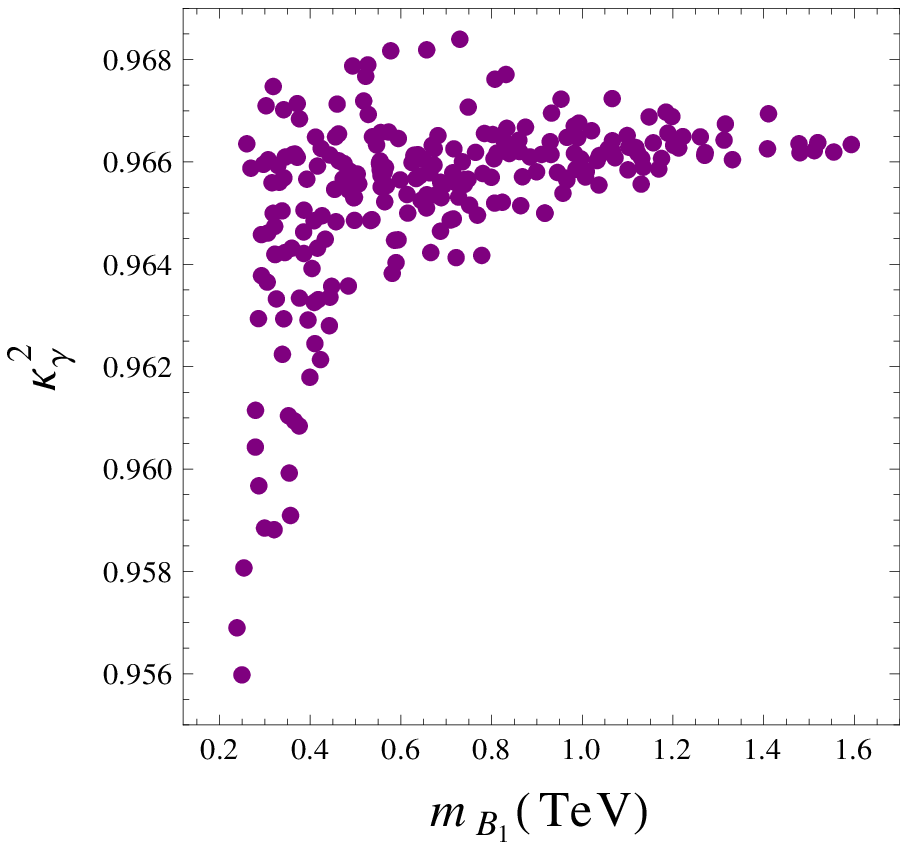}}
\caption{The distributions of $\kappa_\gamma$ values entering eq.~(\ref{R_vs_kappas}) as a function of (a) $m_{T_1}$ and (b) $m_{B_1}$ in the 4DCHM for the benchmark point $f=1$ TeV and $g_*=2$. 
%Notice that the mass regions to the left of the vertical dashed-red lines are excluded by direct searches for $t'$ and $b'$ %states. 
\label{fig:kA}}
\end{figure}

To summarise then, we are  in presence of contrasting effects entering eq.~(\ref{R_vs_kappas}). All $\kappa_i^2$ therein tend to diminish, relative to the SM. {However,  the decrease of $\kappa_H^2$ entering the denominator is bigger than the decrease of the $\kappa_Y^2\times\kappa^2_{Y'}$ product in the numerator, so that the net effect could be the increase of  the event rate in comparison to the SM. This dynamics was indeed shown in Fig.~\ref{fig:R} while its details can be seen in Figs.~\ref{fig:kH}--\ref{fig:kA}}.

{In Fig.~\ref{fig:RA} we investigate these effects further for $R_{\gamma\gamma}$, for which (recall) the largest anomaly is seen, plotted as a function of $m_{T_1}$ and $m_{B_1}$. {$R_{\gamma\gamma}$ values can reach 1.1 for $m_{T_1}$ and $m_{B_1}$ as low as 400 GeV, decreasing to lower values for higher masses.} However, again, should heavy quark masses be allowed to be below 400 GeV or so, the values for $R_{\gamma\gamma}$ could have be rather large, up to 1.2 (or even more). In fact, quite irrespectively of the actual values attained by $R_{\gamma\gamma}$, the tendency in Fig.~\ref{fig:RA} is clear enough.} There is a consistent `leakage' of points towards $R_{\gamma\gamma}>1$, the more so the lighter $m_{T_1}$ and $m_{B_1}$. The relevance of this result is twofold. On the one hand, this call for a thorough re-examination from an experimental point of view of the actual limits on the $t'$ and $b'$ states, especially for low masses, certainly affording an accuracy well beyond the one stemming from the rudimentary approach we have adopted in Figs.~\ref{fig:limit-tp}--\ref{fig:limit-bp}. {On the other hand,  we would like to argue that statements from previous literature, mentioning that accurate predictions can be made in the infinite $t'$ and $b'$  mass limit \cite{Azatov:2011qy,Azatov:2012rd,Azatov:2012ga}, {i.e., those yielding the asymptotic values for large  $m_{T_1}$ and $m_{B_1}$,
may not be applicable to our concrete realisation of the 4DCHM. (We will explicitly show in App.~\ref{appe:b} that the main effects responsible for such
differences are due to the dynamics in the $b$-sector, particularly  in regions of parameter space corresponding to a low mass of the extra fermions).}

\begin{figure}[!t]
\centering
\subfloat[]{
\includegraphics[width=0.45\linewidth]{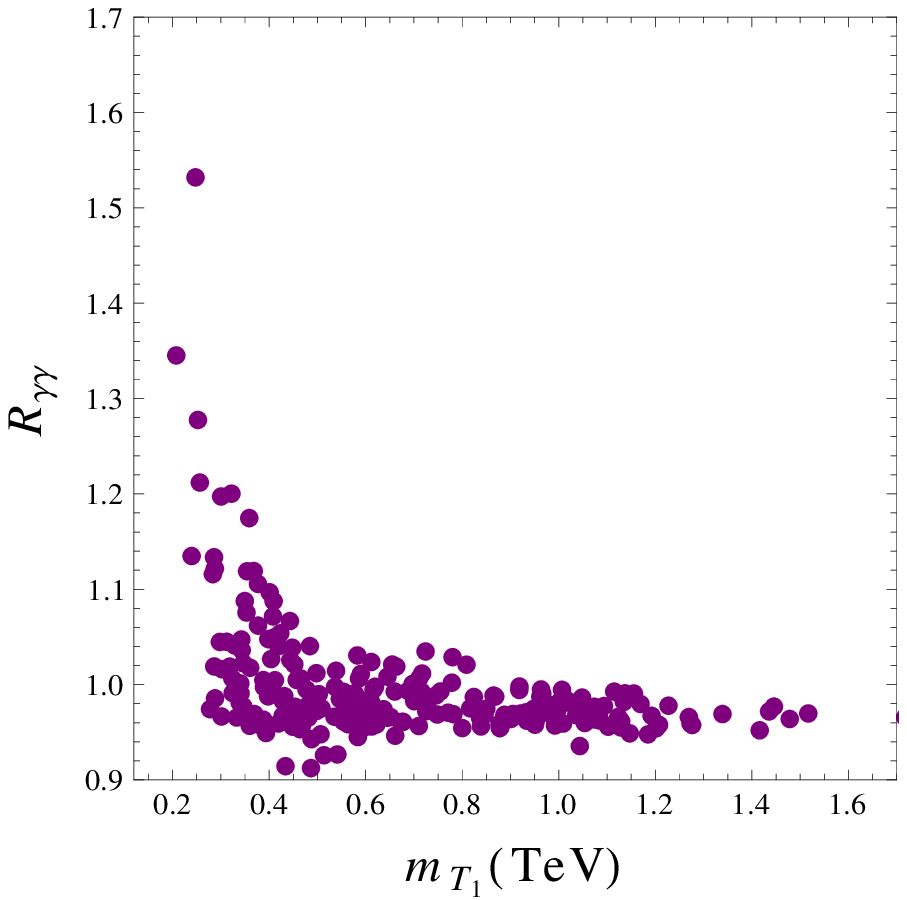}}
\subfloat[]{
\includegraphics[width=0.45\linewidth]{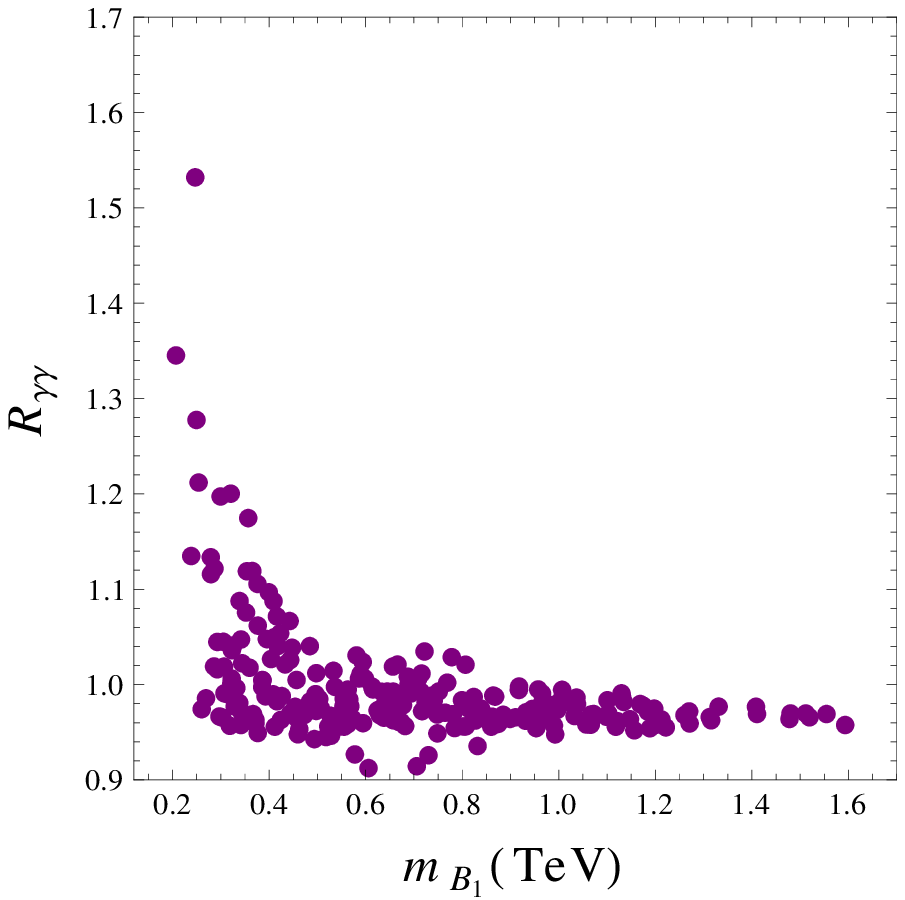}}
\caption{The distributions of $R_{\gamma\gamma}$ values entering eq.~(\ref{newR}) as a function of (a) $m_{T_1}$ and (b) $m_{B_1}$ in the 4DCHM for the benchmark point $f=1$ TeV and $g_*=2$.
%Notice that the mass regions to the left of the vertical dashed-red lines are excluded by direct searches for $t'$ and $b'$ %states. 
\label{fig:RA}}
\end{figure}

However, for the time being, we take the limits on $m_{T_1}$ and $m_{B_1}$ as we obtained them at face value and collect all the results produced, including those for the other $f$ and $g_*$ benchmarks, and compare them to the experimental results of ATLAS~\cite{ATLAS-CONF-2012-170} and CMS~\cite{CMS-PAS-HIG-12-045}  collected in Tab.~\ref{tab:R}.
For each $(f,g_*)$ benchmark we scan over the other free parameters and remove points that do not survive the $t^\prime$, $b^\prime$ and charge 5/3 quark direct search constraints. For the remaining points we calculate $R_{YY}$ for $YY=\gamma\gamma$, $WW$, $ZZ$, $b\bar{b}$. The results are shown in Fig.~\ref{fig:Matt} as a series of normalised histograms in order to demonstrate the number of points in the scan taking particular values of $R_{YY}$ and the full range of values obtained is shown. The experimental measurements for $R_{YY}$ are shown by black and white points with 68\% Confidence Level (CL) error bars. As the scale $f$ is increased, the values of $R_{YY}$ become more sparse { (owing to EWPT constraints)}. This is because parameters in the model become more tightly constrained as this scale grows larger.
%{\bf SM: Matt to replace the current  figure as discussed and provide one paragraph of text commentary}. 
\begin{figure}[!t]
\centering
\includegraphics[width=0.9\linewidth]{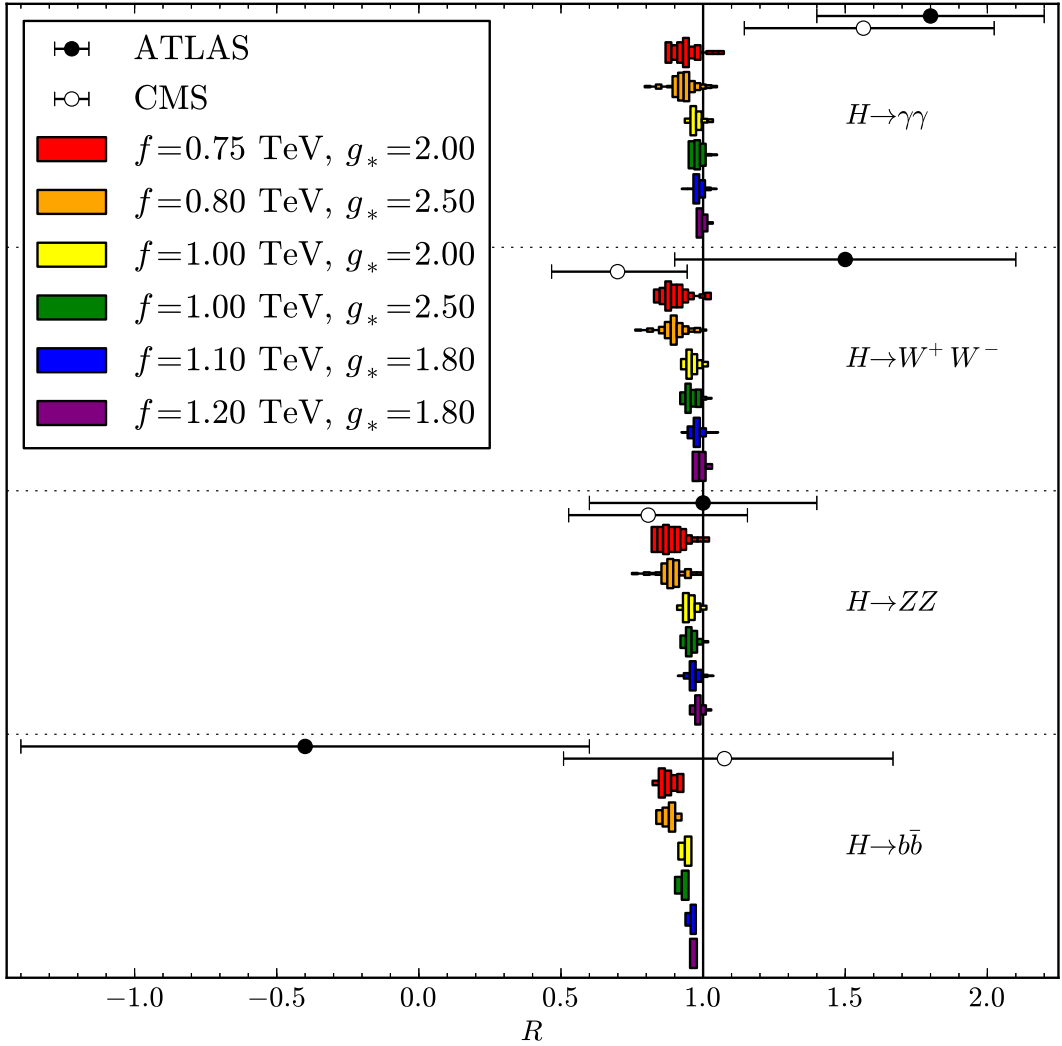}
\caption{Comparison of the $R$'s from eq.~(\ref{newR}) with the measured experimental values by ATLAS~\cite{ATLAS-CONF-2012-170} and CMS~\cite{CMS-PAS-HIG-12-045} (see Tab.~\ref{tab:R})  in the 4DCHM for all benchmark points in $f$ and $g_*$. All points generated here are compliant with direct searches for $t'$s, $b'$s and exotic states with charge 5/3.}
\label{fig:Matt}
\end{figure}
\begin{figure}[!t]
\centering
\includegraphics[width=0.9\linewidth]{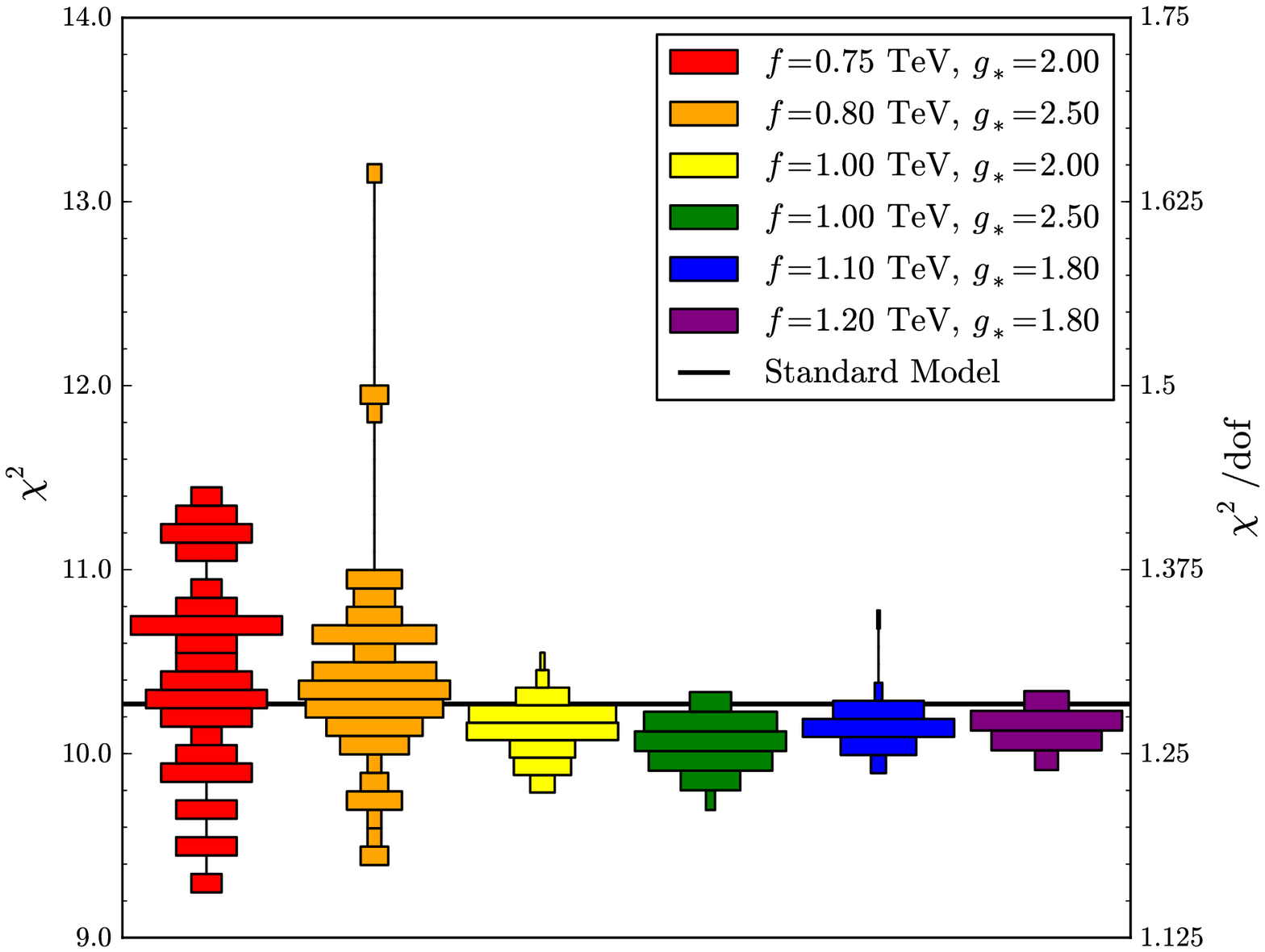}
\caption{The $\chi^2$ fit (as described in the text) in the 4DCHM for all benchmark points in $f$ and $g_*$. All points generated here are compliant with direct searches for $t'$s, $b'$s and exotic states with charge 5/3.}
\label{fig:chi2}
\end{figure}

{In order to have a clear picture on  how the 4DCHM fares against LHC data, particularly in relation to the SM, in a  quantitative way, we calculate the $\chi^2$ goodness of our fit for the ATLAS~\cite{ATLAS-CONF-2012-170} and CMS~\cite{CMS-PAS-HIG-12-045} data from Tab.~\ref{tab:R}}. We assume that all the channels and experiments are independent and simply sum them in the $\chi^2$ function, giving us eight degrees of freedom (dof). The value of $\chi^2$ for each parameter scan point (surviving the experimental constraints discussed above) is shown in Fig.~\ref{fig:chi2} using normalised histograms. The value of our $\chi^2$ function for the SM (i.e., $R_{YY} = 1$) is also plotted as a horizontal black line and the figure makes it clear that the 4DCHM represents a better fit to the data than the SM does for most of the benchmarks considered.

Before closing this section, {we would like to stress that we cannot present the $\chi^2$ analysis using the language of contours in the $\kappa_{\gamma}$-$\kappa_{g}$ plane, as it was done in various experimental and phenomenological papers. The main reason is because, for any given point in the $\kappa_{\gamma}$-$\kappa_{g}$ plane, the other parameters entering the $\chi^2$, for example $\kappa_H$, will be different. Therefore, for each point over the 4DCHM parameter space, there should be a separate corresponding $\chi^2$ contour over the $\kappa_{\gamma}$-$\kappa_{g}$ plane and the emerging description would be rather cumbersome. So, for the concrete realisation of the 4DCHM used here, the approach  $\kappa_H\simeq 1$ chosen in many papers on the subject does not work and the $\chi^2$ results for the corresponding parameter space cannot be presented in an elegant way as a single contour. We believe that this is the case for many classes of models including many from Supersymmetry.} 

%{\bf SM: Daniele \& Matt to prepare Fig. 9 as discussed and comment}

%%%%%%%%%%%%%%%%%%%%%%%%%%%%%%%%%%%%%%%%%%%%%%%%%%%%%%%%%%

\section{Conclusions}
\label{Sec:Conclusions}
\noindent

{In summary, we have shown that the 4DCHM could provide a better explanation  than the SM of  the current LHC data pointing to the discovery of a neutral Higgs boson with mass about  125--126 GeV. This is well  summarised in Figs.~\ref{fig:Matt}--\ref{fig:chi2}.

After systematically scanning the parameter space of the 4DCHM and illustrating its phenomenology for several benchmark points, we have shown that a moderate enhancement  in  the $H\to \gamma\gamma$ channel with respect to the SM predictions is a possible feature of this model and can be as large as about 1.1, somewhat below the central values of the experimental measurements.

However, we have also found that this enhancement could  potentially be larger, realistically up to 1.3, due to the contribution from the  heavy $t'$ and $b'$ fermions of the 4DCHM  with mass just below 400 GeV,  i.e.,  precisely when entering mass regions apparently excluded but for which there are no data from direct searches, only simple extrapolations that we attempted here. So, a thorough re-assessment from the experimental side is required in this respect.

The main source of the enhancement of the $H\to \gamma\gamma$ channel is in the reduction of the $H\to b\bar{b}$ partial width due to $b-b'$ mixing effects which in turn lead to the reduction of the total Higgs boson width and the enhancement of  all decay channels, including the di-photon one. Competing effects emerge though from the (effective) $Hgg$ coupling becoming simultaneously smaller.
 
The reduction  of the total Higgs boson width is a generic feature of not only the composite Higgs model studied here but also of many other classes of BSM theories (including Supersymmetry). So, we would like to encourage the experimental groups to take this fact into account and perform fits of  specific models including the 4DCHM one, where deviations of the total Higgs boson width from the SM one can conveniently be parametrised with the help of few a additional variables. For example, in case of the 4DCHM, this can be done using just one additional parameter describing the $Hb\bar b$ coupling.}

{Finally, a relevant by-product of our analysis has been to show that several approximations adopted in literature within the 4DCHM (or similar scenarios), which essentially make predictions in the limit in which the masses of the heavy fermions (and possibly heavy gauge bosons) are infinitely heavy, cannot generally be accurate over the entire parameter space of the corresponding model, so that they should neither be extrapolated nor be generalised.} 
%{\color{green}Finally, a relevant by-product of our analysis has been to show that there are several interesting effects acting on the couplings $ggH$ and %$H\gamma\gamma$ in a region of the parameter space where the asymptotic approximation usually adopted in literature within similar scenarios does not hold.}

%%%%%%%%%%%%%%%%%%%%%%%%%%%%%%%%%%%%%%%%%%%%%%%%%%%%%%%%%%

\section{Note added}
\label{Sec:Noted}
\noindent
After the completion of our work, both ATLAS and CMS reported new results at the Moriond conference. However, we would like to stress that their impact does not change qualitatively our main conclusion, i.e., that the 4DCHM appears to be a viable solution to the LHC Higgs data and provides substantial regions of its parameter space yielding a better fit to the latter than the SM. In particular, the $R_{\gamma\gamma}$ rate measured by the CMS collaboration has substantially lowered, which has rendered the 4DCHM parameter space with slightly higher values of $f$ ($f \simeq 0.8$ versus $0.75$ for the pre-Moriond LHC data) somewhat more preferred by the fit \cite{Barducci:2013ica}.
 
%%%%%%%%%%%%%%%%%%%%%%%%%%%%%%%%%%%%%%%%%%%%%%%%%%%%%%%%%%

\section*{Acknowledgements} 
\noindent

GMP would like to thank Guido Macorini and Dominik St\"okinger for helpful discussions. He is also grateful to the INFN (Sezione di Cagliari) for the hospitality and logistical support during the completion of this work. DB, AB and SM are financed in part through the NExT Institute. The work of GMP has been supported by the German Research Foundation DFG through Grant No.\ STO876/2-1, by BMBF Grant No.\ 05H09ODE and by the European Community's Seventh Framework Programme (FP7/2007-2013) under grant agreement n.~290605 (PSI-FELLOW/COFUND). DB, SDC and GMP would like to thank the Galileo Galilei Institute (GGI) in Florence for hospitality while part of this work was carried out.

\clearpage

%%%%%%%%%%%%%%%%%%%%%%%%%%%%%%%%%%%%%%%%%%%%%%%%%%%%%%%%%%%%%
\appendix
\section{Loop-induced $H$ decay modes} 
\label{appe:a}
\noindent

In this appendix we analyse the loop-induced Higgs decay modes that we have considered in detail in the main body of the paper, i.e. the EW one, $\gamma\gamma$, as well as the QCD one, $gg$\footnote{{In addition, we have also computed the $H\to Z\gamma $ case. However, since it has little phenomenological relevance here and the corresponding formulae are rather complicated, we will not discuss it in detail.}}. We make use of the Passarino-Veltman decomposition following the notation of \cite{Denner:1991kt}. In the absence of charged scalars, the only involved topologies contain (massive) fermions and charged gauge vector bosons in the loop, i.e., $b$s, $t$s (and, possibly, $W$s) in the SM plus $b'$s, $t'$s (and, possibly, $W'$s) in the 4DCHM. Since the loop structures are the same for all fermions and bosons, but different one from the other, in the following we consider only the case of $b$s and $W$s as representative example of the two species, respectively. In Tab.~\ref{tab:feyn} we collect the whole set of Feynman rules which are necessary to evaluate the bosonic and fermionic loop-induced Higgs decay modes in the 4DCHM. {Finally, we note that we have included NLO QCD corrections in our implementation of the $Hgg$ vertex but also remark that they are model
independent (as the same particle species flow in the loops of both the 4DCHM and the SM), so that they always cancel out in the estimate of the $R$ parameters.}

\begin{table}[!htb]
\begin{center}
\begin{tabular}{|l|l|l|} \hline
Fields in the vertex & Factor & Tensorial structure \\ \hline
$\bar{b}_{i}$ \phantom{-} ${b}_{i}$ \phantom{-} ${H}_{}$ \phantom{-}  &
$C_{H b_i\bar{b}_i}$ & 1 \\
$\bar{b}_{i}$ \phantom{-} ${b}_{i}$ \phantom{-} $\gamma^\mu$ \phantom{-}  &
$C_{\gamma b\bar{b}}$ & $\gamma^\mu$ \\
$\bar{b}_{i}$ \phantom{-} ${b}_{j}$ \phantom{-} ${H}_{}$ \phantom{-}  &
$C_{H b_j\bar{b}_i}$ & $R_{Hb_j\bar{b}_i} (1+\gamma^5)+L_{Hb_j\bar{b}_i} (1-\gamma^5)$ \\
$\bar{b}_{i}$ \phantom{-} ${b}_{j}$ \phantom{-} ${Z}^\mu$ \phantom{-}  &
$C_{Z b_j\bar{b}_i}$ & $L_{Zb_j\bar{b}_i}\gamma^\mu (1-\gamma^5)+R_{Zb_j\bar{b}_i}\gamma^\mu (1+\gamma^5)$ \\
$\left({W}^{+}_{i}\right)^\mu$ \phantom{-} $\left({W}^{-}_{j}\right)^\nu$ \phantom{-}  ${H}_{}$ \phantom{-} &
$C_{H W_i W_j}$ & $g^{\mu\nu}$ \\
$\left({W}^{+}_{i}\right)^\mu$ \phantom{-} $\left({W}^{-}_{i}\right)^\nu$ \phantom{-} $\gamma^\rho$ \phantom{-}  &
$C_{\gamma WW}$ &  $(p_{W^-}-p_{W^+})g^{\mu\nu}+(p_{\gamma}-p_{W^-})g^{\nu\rho}+(p_{W^+}-p_{\gamma})g^{\rho\mu}$\\
$\left({W}^{+}_{i}\right)^\mu$ \phantom{-} $\left({W}^{-}_{j}\right)^\nu$ \phantom{-} $Z^\rho$ \phantom{-}  &
$C_{Z W_i W_j}$ &  $(p_{W^-}-p_{W^+})g^{\mu\nu}+(p_{Z}-p_{W^-})g^{\nu\rho}+(p_{W^+}-p_{Z})g^{\rho\mu}$\\
$\left({W}^{+}_{i}\right)^\mu$ \phantom{-} $\left({W}^{-}_{i}\right)^\nu$ \phantom{-} $\gamma^\rho$ \phantom{-} $\gamma^\sigma$ &
$A_{\gamma\gamma WW}$ &  $2g^{\mu\nu}g^{\rho \sigma}-g^{\mu\rho}g^{\nu\sigma}-g^{\mu\sigma}g^{\nu\rho}$ \\
$\left({W}^{+}_{i}\right)^\mu$ \phantom{-} $\left({W}^{-}_{j}\right)^\nu$ \phantom{-} $\gamma^\rho$ \phantom{-} $Z^\sigma$ &
$A_{\gamma Z W_{i}W_{j}}$ &  $2 g^{\mu\nu}g^{\rho \sigma}-g^{\mu\rho}g^{\nu\sigma}- g^{\mu\sigma}g^{\nu\rho}$ \\
 \hline
\end{tabular}
\end{center}
\caption{Relevant Feynman rules for the evaluation of loop-induced Higgs decay modes via $b$s and $W$s in a 4DCHM.\label{tab:feyn}}
\end{table}

We summarise the properties among coefficients which are useful for our aim\footnote{Notice that the case $i\ne j$ is only relevant for the $H\to Z\gamma$ decay (see previous footnote).}:
\begin{eqnarray}
C_{H b_j\bar{b}_i}=C_{H b_i\bar{b}_j}, \quad & R_{Hb_j\bar{b}_i}=L_{Hb_i\bar{b}_j}, \quad & \  \nonumber \\
C_{Z b_j\bar{b}_i}=C_{Z b_i\bar{b}_j}, \quad & R_{Zb_j\bar{b}_i}=R_{Zb_i\bar{b}_j}, \quad & L_{Zb_j\bar{b}_i}=L_{Zb_i\bar{b}_j},\nonumber \\
C_{H W_j W_i}=C_{H W_i W_j}, \quad & A_{\gamma\gamma WW}=-C_{\gamma WW}^2, \quad & C_{Z W_j W_i}=C_{Z W_i W_j}, \nonumber \\
A_{\gamma Z W_{i}W_{j}}=A_{\gamma Z W_{j}W_{i}}, \quad & A_{\gamma Z W_{i}W_{j}}=-2C_{Z W_i W_j}C_{\gamma WW}, \quad & \ 
\end{eqnarray}
for which the following relations hold
\begin{eqnarray}
C_{\gamma b\bar{b}}=eQ_{b},\quad & R_{Hb_i\bar{b}_i}=L_{Hb_i\bar{b}_i}=\frac{1}{2}, \nonumber \\
C_{\gamma WW}=e,\quad & A_{\gamma\gamma WW}=-e^2. 
\end{eqnarray}

\subsection{$H\to \gamma \gamma$ via $b$s and $W$s}

The contributions to the amplitude of the process $H\to \gamma \gamma$ are only due to diagonal couplings between Higgs and photons to $b_i$, $t_i$ and $W_j$ (with $i=0,\dots,8$ and $j=0,\dots,3$ extended to both the SM, here labelled with $i=j=0$, and 4DCHM states, see Tab.~\ref{table:partspec}). Both the spinor- and vector-induced structures of the effective Higgs-to-photons coupling are well-established in literature \cite{Shifman:1979eb}, hence the only difference with respect to the SM case lies in the multiplicity and mass spectra of the particle content of the model.

The fermionic contribution to the amplitude (as said we consider the contribution from the $b$s as representative) is:
\begin{eqnarray} 
\mathcal{M}_b(H\to \gamma \gamma)=i\sum_{i=0}^{8}F_{b_i}\left(2+4M_{b_i}^2C_0-M_H^2C_0\right)\left(g^{\mu \nu}k_1\cdot k_2-k_1^\nu k_2^\mu\right)\varepsilon^*_\mu(k_1)\varepsilon^*_\nu(k_2),
\end{eqnarray}
where
\begin{eqnarray}
C_0&=&C_0(0,0,M_H^2,M_{b_i}^2,M_{b_i}^2,M_{b_i}^2), \\
F_{b_i}&=&\frac{N_c M_{b_i}}{2\pi^2M_H^2} C_{\gamma b\bar{b}}^2 C_{H b_i\bar{b}_i}.
\end{eqnarray}
 In Fig.~\ref{FD_HAA_F}  we show the fermionic contributions to the amplitude ${\mathcal M} (H\to \gamma \gamma)$
in terms of Feynman diagrams.

The charged vector contribution to the amplitude is:
\begin{eqnarray}
&\ &\mathcal{M}_W(H\to \gamma \gamma)=\nonumber \\
&=&i \sum_{i=0}^{3}F_{W_i}\left(6+\frac{M_H^2}{M_{W_i}^2} + 12 C_0 M_{W_i}^2 - 6 C_0 M_H^2 \right)\left(g^{\mu \nu}k_1\cdot k_2-k_1^\nu k_2^\mu\right)\varepsilon^*_\mu(k_1)\varepsilon^*_\nu(k_2),
\end{eqnarray}
where
\begin{eqnarray}
C_0&=&C_0(0,0,M_H^2,M_{W_i}^2,M_{W_i}^2,M_{W_i}^2), \\
F_{W_i}&=&\frac{1}{8 \pi^2 M_H^2} C_{\gamma WW}^2 C_{H W_i W_i}.
\end{eqnarray}
In Fig.~\ref{FD_HAA_W}  we show the bosonic contributions to the amplitude ${\mathcal M}(H\to \gamma \gamma)$ in terms of Feynman diagrams.

\subsection{$H \to gg$ via $b$s}

The contributions to the amplitude of the process $H\to gg$ are only due to diagonal couplings between Higgs and gluons to $b_i$, $t_i$ (with $i=0,..,8$). The spinor-induced structure of the effective gluon-fusion coupling is well-established in literature \cite{Wilczek:1977zn,Georgi:1977gs,Ellis:1979jy,Rizzo:1979mf}. The width $\Gamma(H \to gg)$ is obtained, in this case, as a trivial modification of the formulae for the width of $H\to \gamma \gamma$ where the contributions from $W$s has been stripped off and only the ones from (coloured) fermions are considered, i.e., $\Gamma_f(H \to \gamma \gamma)$:
\begin{eqnarray}
\Gamma(H \to gg) \longrightarrow \frac{2g_S^4}{N_c^2e^4Q_f^4}\Gamma_f(H \to \gamma \gamma).
\end{eqnarray}

\begin{figure}[!htb]
\subfloat[]{\label{FD_HAA_F_1}
  \includegraphics[width=0.32\textwidth]{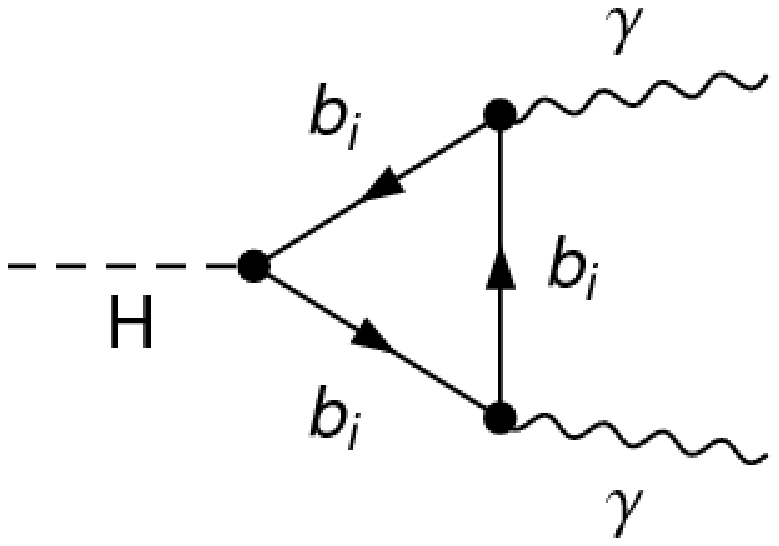}}
\subfloat[]{\label{FD_HAA_F_2}
  \includegraphics[width=0.32\textwidth]{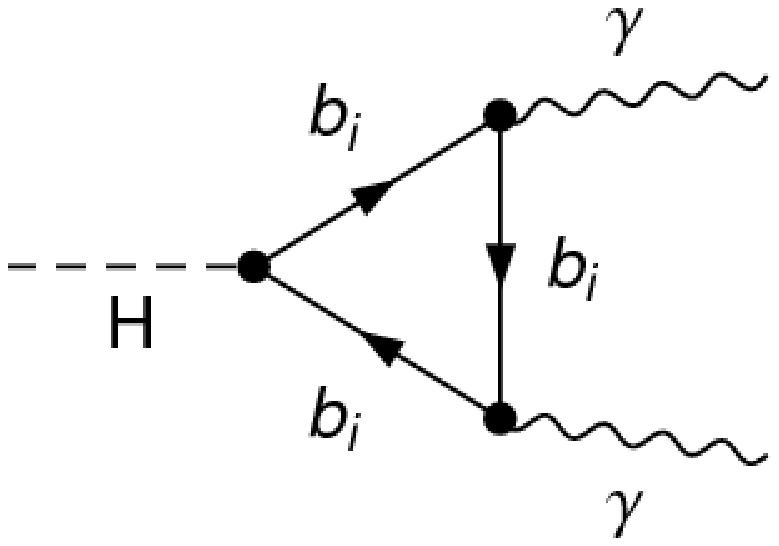}}
\caption{$H\to \gamma \gamma$ induced by a charged fermion loop. \label{FD_HAA_F}}
\end{figure}

\begin{figure}[!htb]
\subfloat[]{\label{FD_HAA_W_1}
  \includegraphics[width=0.32\textwidth ]{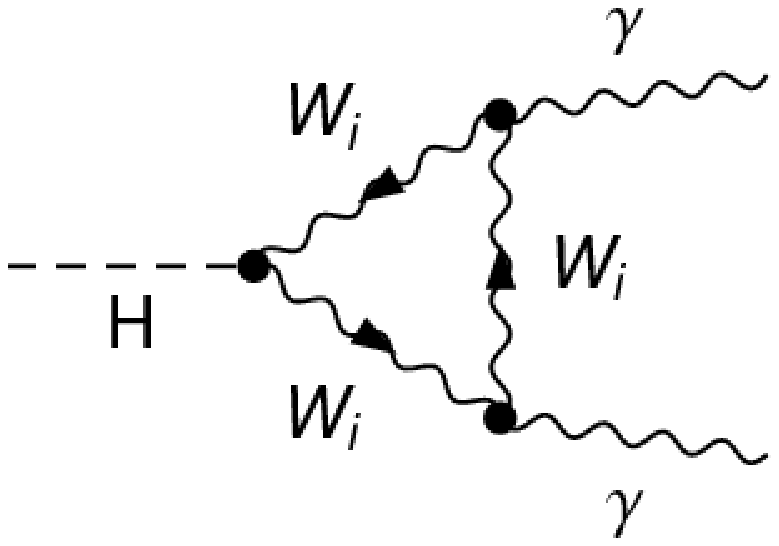}}
\subfloat[]{\label{FD_HAA_W_2}
  \includegraphics[width=0.32\textwidth ]{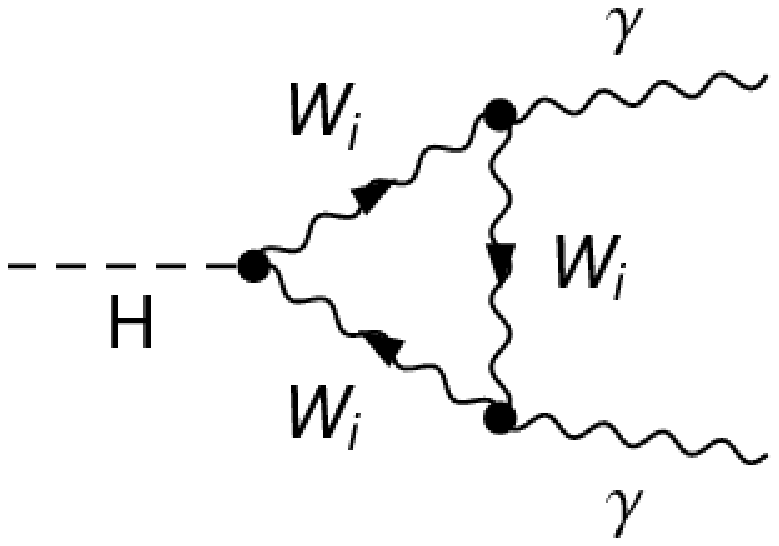}}
\subfloat[]{\label{FD_HAA_W_3}
  \includegraphics[width=0.32\textwidth ]{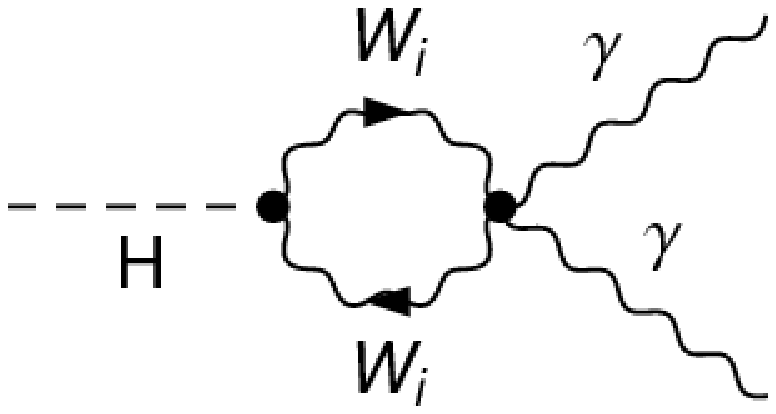}}
\caption{$H\to \gamma \gamma$ induced by a charged vector loop. \label{FD_HAA_W}}
\end{figure}

%\clearpage

%%%%%%%%%%%%%%%%%%%%%%%%%%%%%%%%%%%%%%%%%%%%%%%%%%%%%%%%%%

\clearpage

%%%%%%%%%%%%%%%%%%%%%%%%%%%%%%%%%%%%%%%%%%%%%%%%%%%%%%%%%%
\section{Loops of heavy fermions} 
\label{appe:b}
\noindent

In this appendix, we study closely the Higgs couplings to both the SM and 4DCHM particles entering the production and decay processes studied in the main body of the paper. In particular, we will illustrate  how such couplings interplay with the loop functions already introduced and thus affect the dynamics entering $H\to gg$ and $H\to\gamma\gamma$.

\subsection{Higgs couplings in the 4DCHM}

In order to analyse the various contributions to the Higgs production and decay rates and in particular their interplay inside the loops, it is useful to show the behaviour of the most relevant Higgs  couplings to the gauge bosons, $g_{HV_iV_i}$, and fermions, $g_{Hf \bar f }$ (where $V_i$ is a label identifying both the SM ($W$ and $Z$) and 4DCHM ($W'$ and $Z'$) gauge bosons and $f$ corresponds to both SM ($t$ and $b$) and 4DCHM ($t'$ and $b'$) quarks), in terms of the following model parameters: $f$ (the compositeness scale), $g_*$ (the coupling in the new gauge sector), $\Delta_{t,b/L,R}$  (the mixing parameters between the elementary and composite sectors), $Y_{T,B}$ and $M_{Y_{T,B}}$  (the Yukawa parameters of the composite sector) and $m_*$  (the mass parameter of the new  resonances in the fermionic sector).  

In Fig.~\ref{HWW} we plot the ratios of the couplings of the Higgs boson to the charged gauge bosons in the 4DCHM, $g_{HWW}$, $g_{HW_2W_2}$ and $g_{HW_3W_3}$ ($W_1$ is not coupled to the Higgs boson).
 \begin{figure}[htb]
\centering
\subfloat[]{
\includegraphics[width=0.4\linewidth]{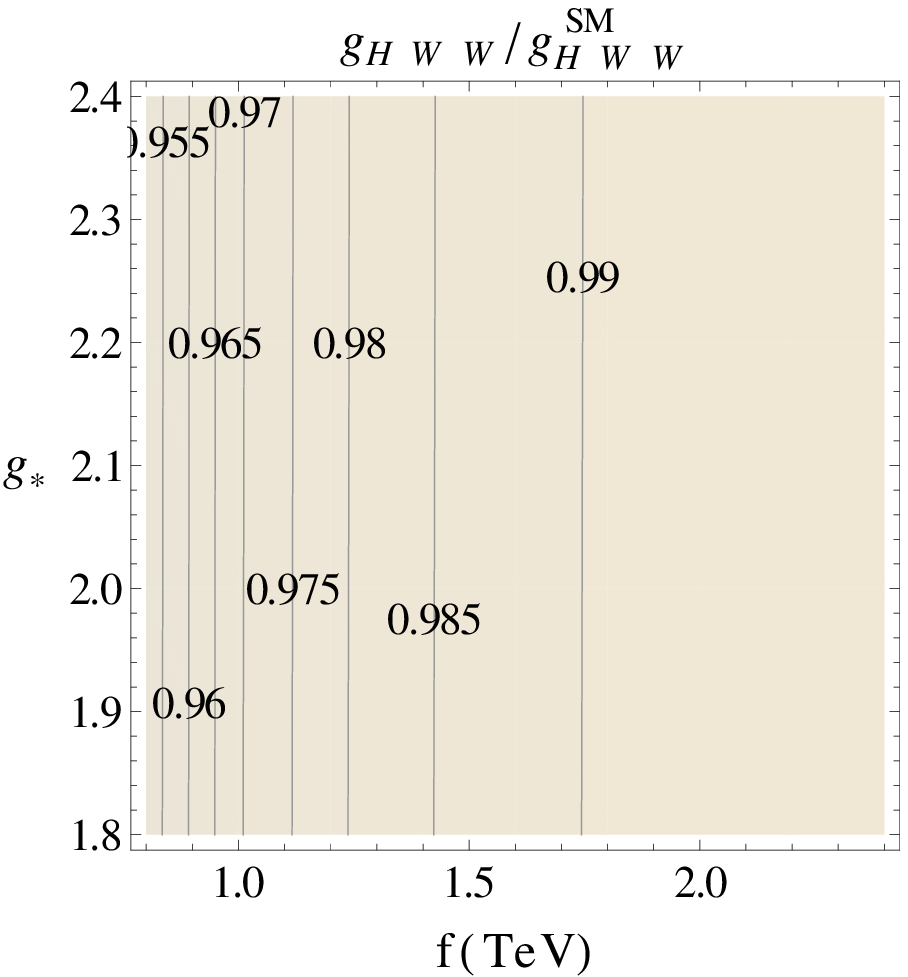}}\\
\subfloat[]{
\includegraphics[width=0.4\linewidth]{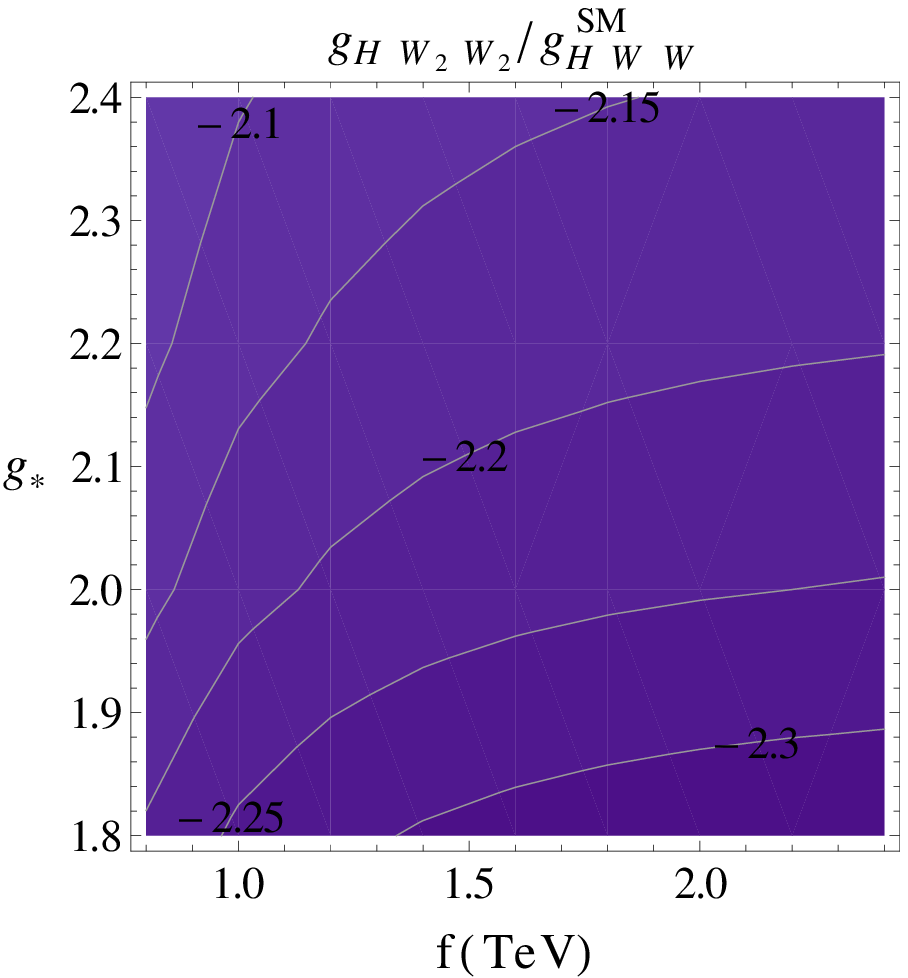}}
\subfloat[]{
\includegraphics[width=0.4\linewidth]{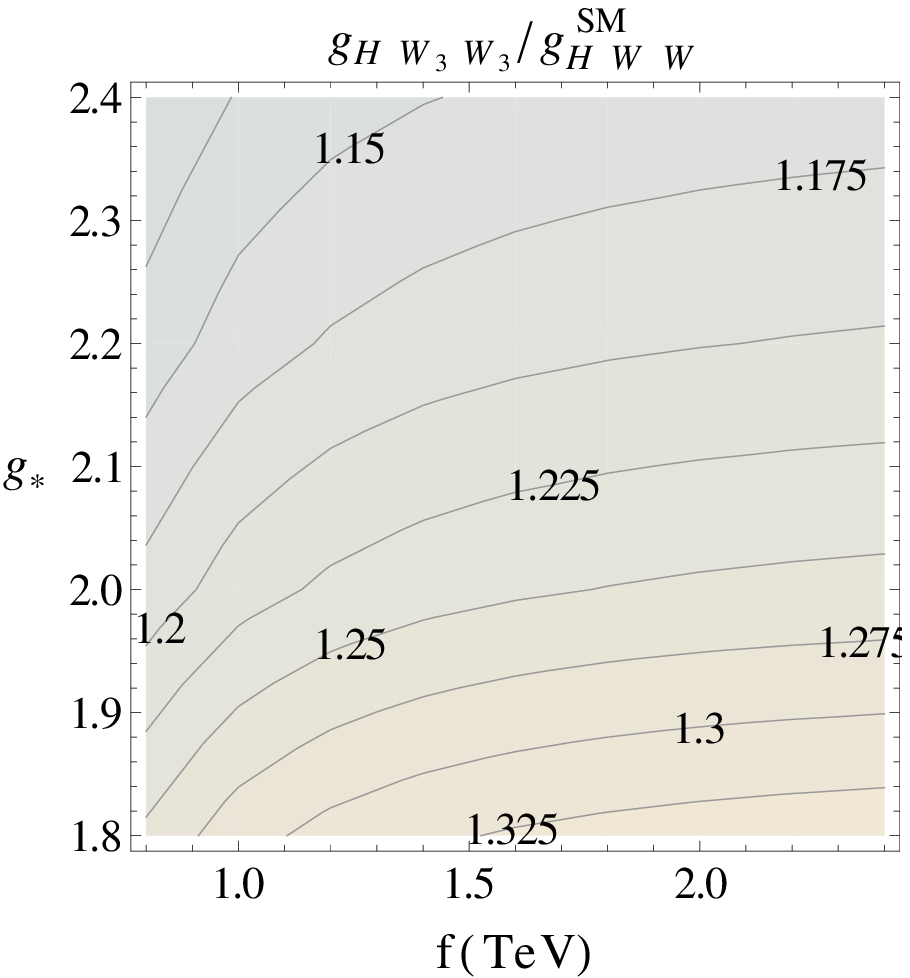}}
\caption{Couplings of the Higgs boson in the 4DCHM to charged gauge boson pairs ($W$ top panel, $W_2$ bottom-left panel, $W_3$ bottom-right panel) normalised to the SM value. \label{HWW}}
\end{figure}

As it is clear from the figure, the value of the ${HWW}$  coupling in the 4DCHM  is slightly modified with respect to the SM one, by a negative 1-2\% correction. In contrast, as expected, the extra gauge bosons can be {sizably} coupled to the composite Higgs state. In particular, the latter can have alternate signs. This fact can play a role in the cancellations among different gauge boson contributions that could occur in the $H\to\gamma\gamma$ loops, albeit affecting the phenomenology very little, given the values of the 4DCHM gauge boson masses\footnote{Notice that the number of couplings entering the $H\to \gamma Z$ loops is much higher, as also non-diagonal, i.e.,  $g_{HW_iW_j}$ with $i\ne j$ contributions are allowed. However, this decay channel is both currently inaccessible at the LHC and also plays a very suppressed role in the definition of the total Higgs width, so that we will not dwell further on these couplings here.}. A rather similar situation holds for the couplings of the Higgs to the neutral gauge bosons of the 4DCHM, i.e., $Z$, $Z_2$, $Z_3$ and $Z_5$, as shown in Fig.~\ref{HZZ} ($Z_1$ and $Z_4$ are not coupled to the Higgs boson). Typical couplings for the neutral gauge bosons are similar to those for the charged ones.

 \begin{figure}[htb]
\centering
\subfloat[]{
\includegraphics[width=0.4\linewidth]{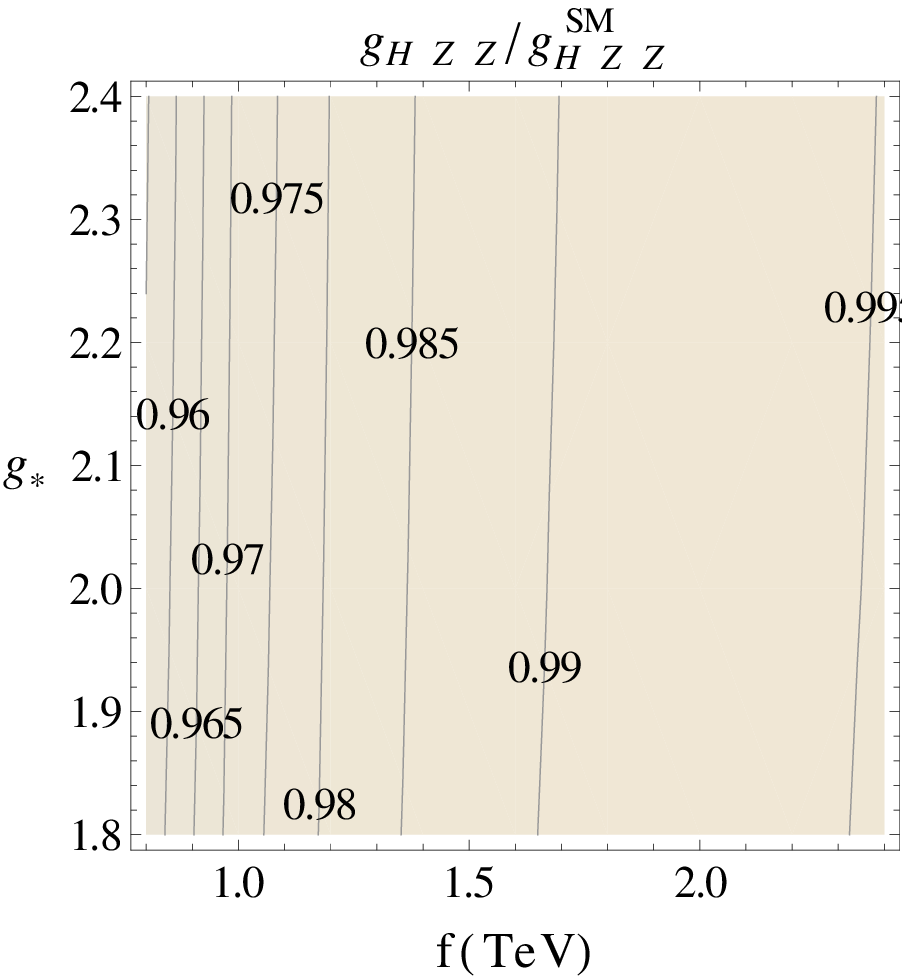}}
\subfloat[]{
\includegraphics[width=0.4\linewidth]{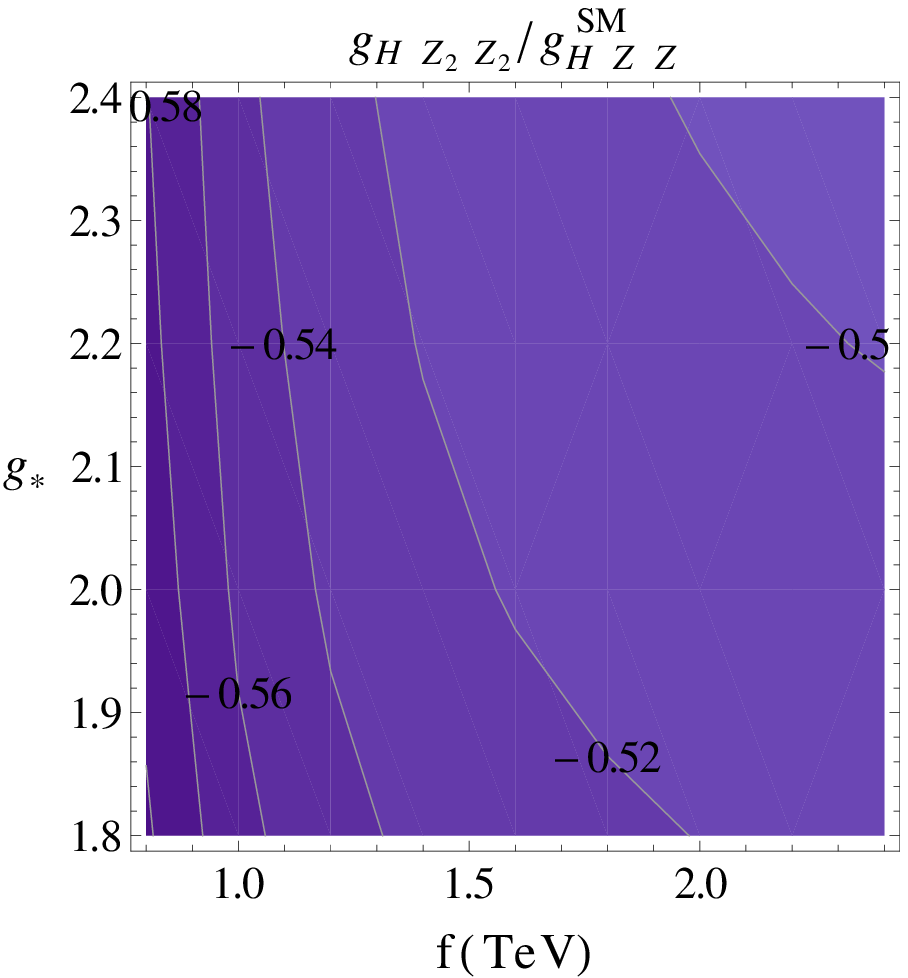}}\\
\subfloat[]{
\includegraphics[width=0.4\linewidth]{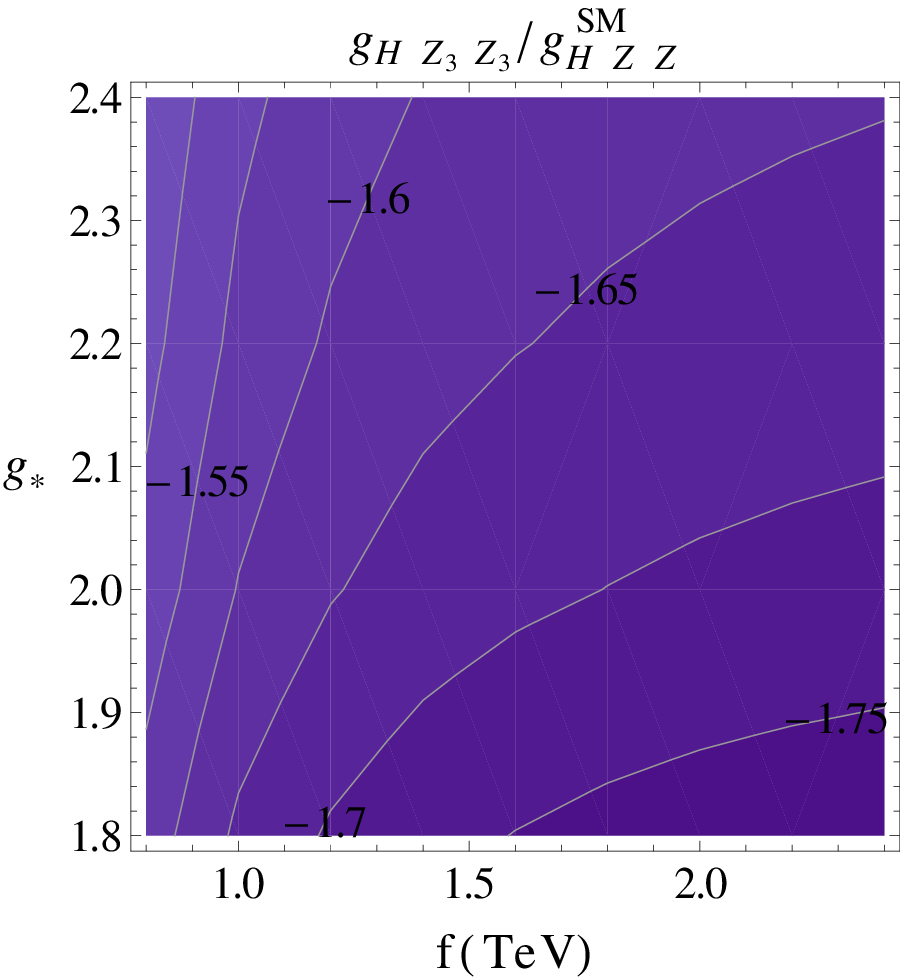}}
\subfloat[]{
\includegraphics[width=0.4\linewidth]{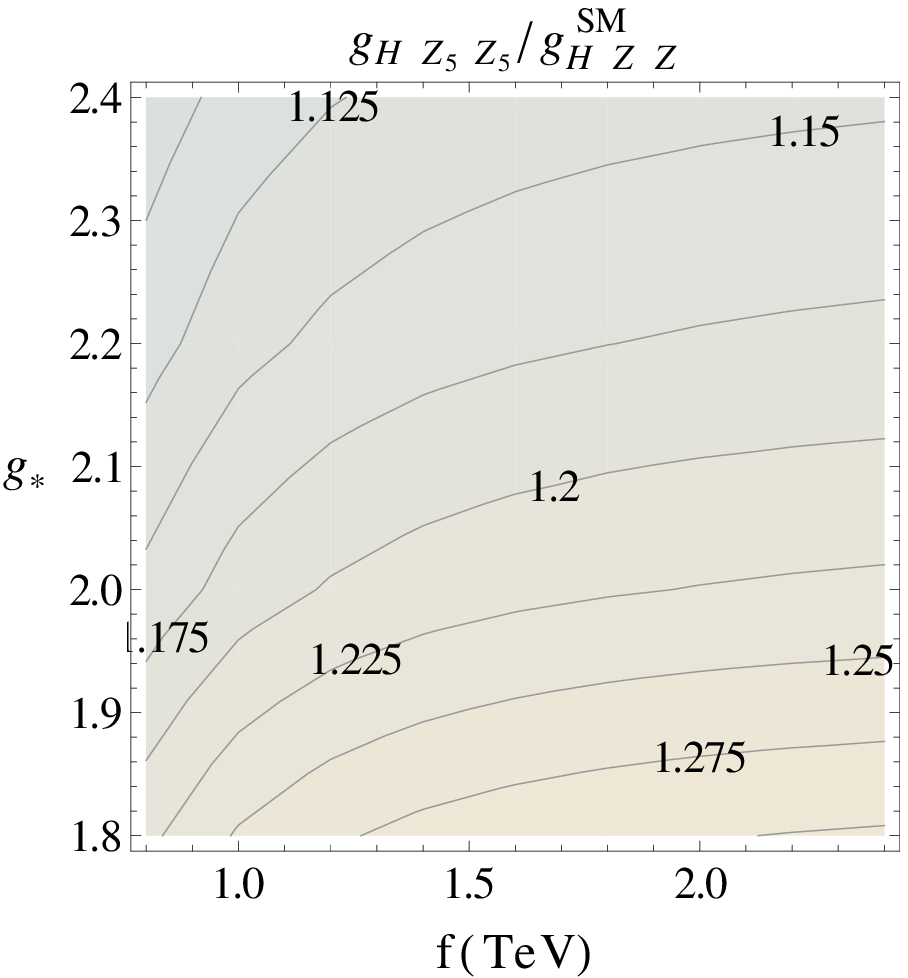}}
\caption{Couplings of the Higgs boson in the 4DCHM to neutral gauge boson pairs ($Z$ top-left  panel, $Z_2$ top-right  panel, $Z_3$ bottom-left panel, $Z_5$ bottom-right panel) normalised to the SM value. \label{HZZ}}
\end{figure}
In the fermion sector, the 4DCHM Higgs couplings depend on other parameters than just $f$ and $g_*$, as recalled above. Hence, it is not possible to present a set of plots similar to the case of the gauge sector. As explained in the main text, we have performed a scan over these additional parameters, and thus present the results in this case as scatter plots. This is done in Fig.~\ref{Hqq} for the case of top and bottom quarks, again   with respect to the SM values for the corresponding flavours, e.g.,  as a function of $m_{T_1}$ and $m_{B_1}$ (the lightest heavy fermion partners of either flavour, respectively)\footnote{Notice that we use here a different benchmark with respect to the main text, for reasons to be explained in the next subsection. }.

\begin{figure}[htb]
\centering
\subfloat[]{
\includegraphics[width=0.4\linewidth]{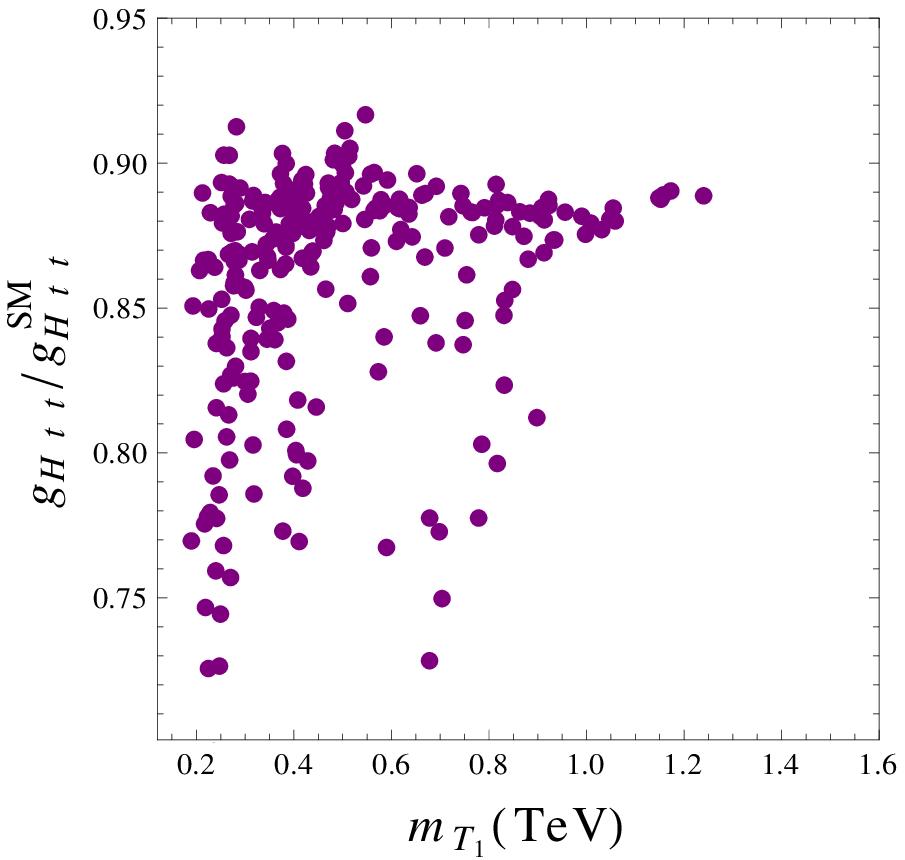}}
\subfloat[]{
\includegraphics[width=0.4\linewidth]{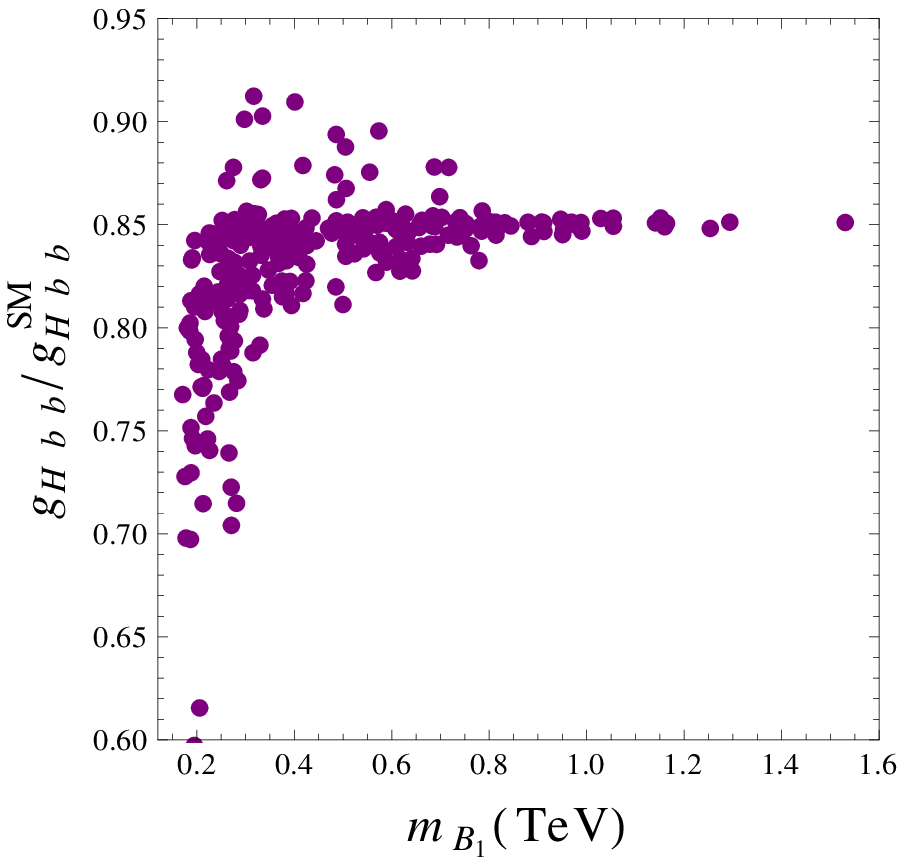}\label{Hqq_b}}
\caption{Couplings of the Higgs boson in the 4DCHM to top (left panel) and bottom ({right} panel) quark pairs normalised to the corresponding SM values, as a function of $m_{T_1}$ and $m_{B_1}$, respectively. Here, $f=0.8$ TeV and $g_*=2.5$. \label{Hqq}}
\end{figure}
As expected, the Higgs couplings to the top and bottom quark pairs are generally smaller than the SM ones. Apart from a region of small $m_{B_1}$, which is apparently experimentally excluded (as shown in Sect.~\ref{Sec:strategy}), $g_{Hbb}$ is around 10$\%$ smaller than the corresponding SM value. This is the main reason  for the reduction of the total width of the Higgs state of the 4DCHM.  Also the coupling of the Higgs to the top quark pairs  is generally smaller than the SM one, by a similar amount, though this is inconsequential to the value of $\Gamma_{\rm tot}(H)|_{\rm 4DCHM}$. Both such couplings though  interplay in both the production and decay loops.

In Fig.~\ref{HQ1Q1} we plot the ratios of the couplings of the Higgs boson of the 4DCHM to the lightest heavy fermion partners in the top and bottom sector with respect to the $g_{Htt}^{\rm SM}$ (as a function of $m_{T_1}$) and $g_{Hbb}^{\rm SM}$  (as a function of $m_{B_1}$) values, respectively. Recalling that the $HT_1T_1$ and $HB_1B_1$ couplings are not of Yukawa type (i.e., they are not proportional to $m_{T_1}$ and $m_{B_1}$, respectively) and that the top Yukawa coupling in the SM is  about 35 times larger than the bottom one, it is clear that this plot highlights the greater importance of $T_1$ couplings relatively to the $B_1$ ones (in fact, this is true for the whole families of $t'$s and $b'$s, respectively, a point to which we will come back later). Notice that the 4DCHM couplings can have opposite signs relatively to the corresponding SM counterparts, hence inducing the possibility of destructive interference with the SM fermion contributions in the production and decay loops.
\begin{figure}[h!]
\centering
\subfloat[]{
\includegraphics[width=0.4\linewidth]{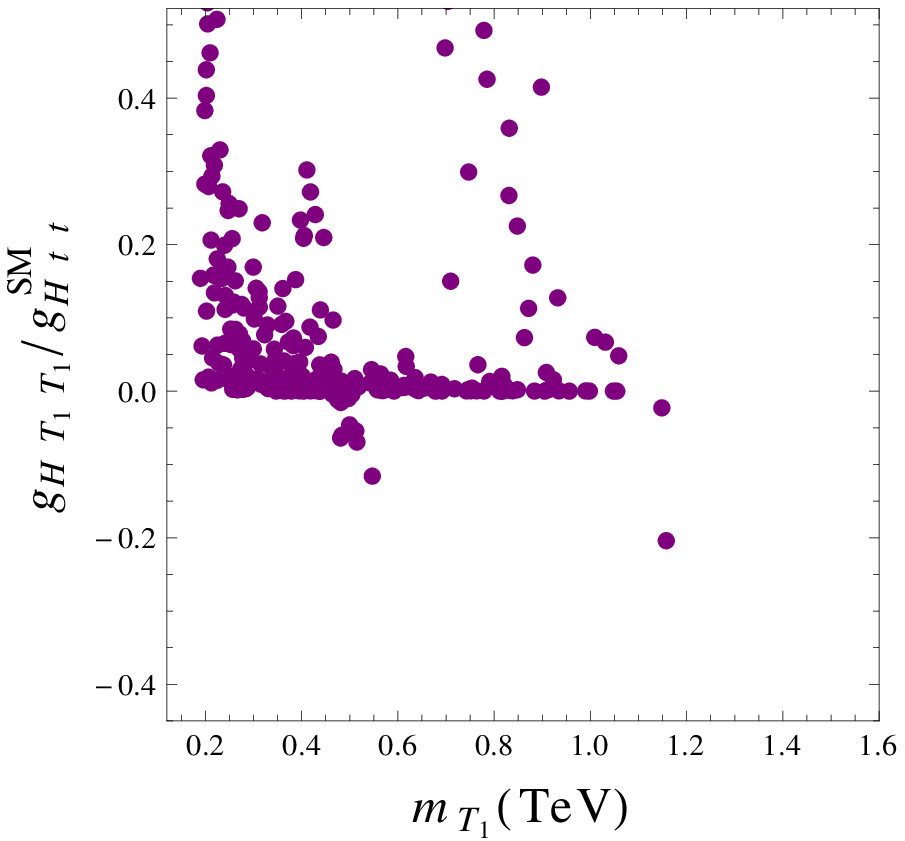}}
\subfloat[]{
\includegraphics[width=0.4\linewidth]{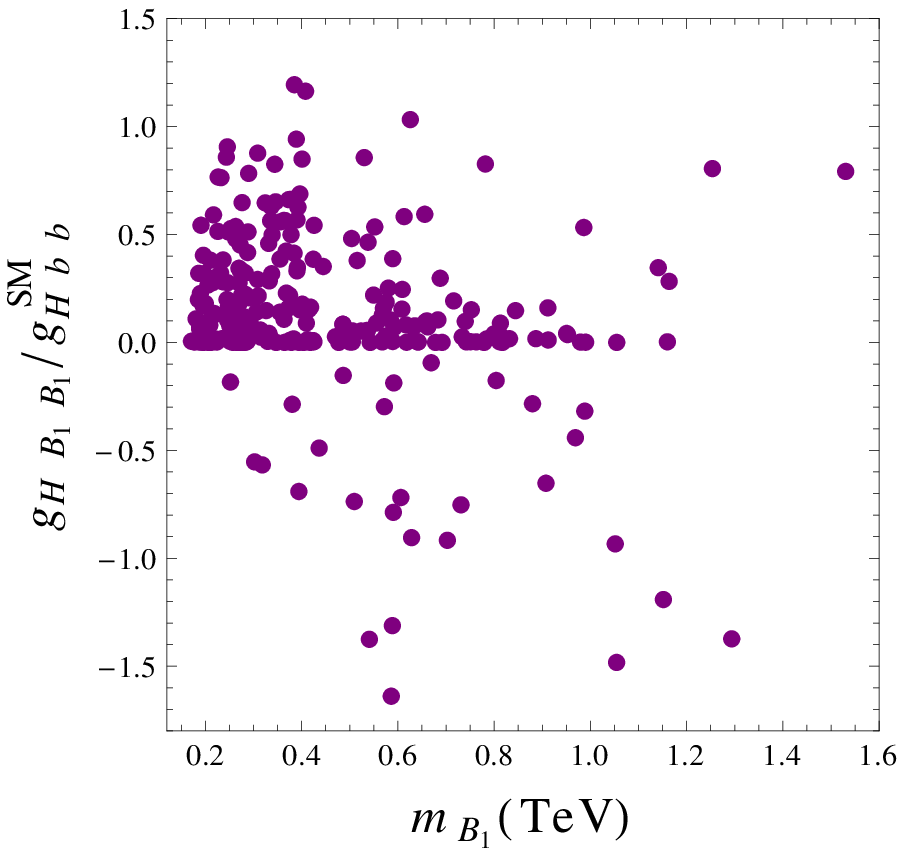}}
\caption{Couplings of the Higgs boson in the 4DCHM to the lightest heavy fermion pairs in the top  (left panel) and bottom (right panel) sector normalised with respect to  the corresponding SM values, 
as a function of $m_{T_1}$ and $m_{B_1}$, respectively. Here, $f=0.8$ TeV and $g_*=2.5$. \label{HQ1Q1}}
\end{figure}

%In the light of the last result, it is of some relevance to also study the couplings of the $T_2$, $T_3$ and $T_4$ states to the Higgs boson of the 4DCHM. This is done in Fig.~\ref{HQ2Q2}  as a function of $m_{T_1}$, so as to compare directly with the $T_1$ case. It is noticeable that all  couplings $g_{HT_2T_2}$, $g_{HT_3T_3}$ and $g_{HT_4T_4}$ are as large as $g_{H T_1T_1}$, with those of the second(third) lightest state having generally the same(opposite) sign as(to) that of the lightest one. 
%\begin{figure}[h!]
%\centering
%\subfloat[]{
%\includegraphics[width=0.4\linewidth]{t2.eps}}\\
%\subfloat[]{
%\includegraphics[width=0.4\linewidth]{t3.eps}}
%\subfloat[]{
%\includegraphics[width=0.4\linewidth]{t4.eps}}
%\caption{Couplings of the Higgs boson in the 4DCHM to second (top panel), third (bottom-left panel) and fourth (bottom-right panel) lightest  heavy fermion pairs  in the top sector with respect to  the corresponding SM values, as a function of $m_{T_1}$ and $m_{B_1}$. Here, $f=0.8$ TeV and $g_*=2.5$. \label{HQ2Q2}}
%\end{figure}

\subsection{$H\to$ $gg$ and $\gamma\gamma$: an overview of the 4DCHM contributions}

In this subsection we describe in some detail the one-loop contributions to the Higgs-to-gluon and Higgs-to-photon amplitudes. Firstly, we collect the various contributions in two main categories: the SM-like ones (from $t$, $b$ and $W$) and the 4DCHM ones (from $T_i$, $B_i$ and $W_j$ with $i=1,\dots ,8$ and $j=2,3$). Secondly, we proceed to establish which are the most sizable ones. In order to evaluate the impact of the extra contributions, we here focus on a benchmark at lower scales, namely $f=0.8$ TeV and $g_*=2.5$. This benchmark is chosen as representative of the typical phenomenology emerging in the 4DCHM. 

\begin{figure}[!ht]
\centering
\subfloat[]{
\includegraphics[width=0.45\linewidth]{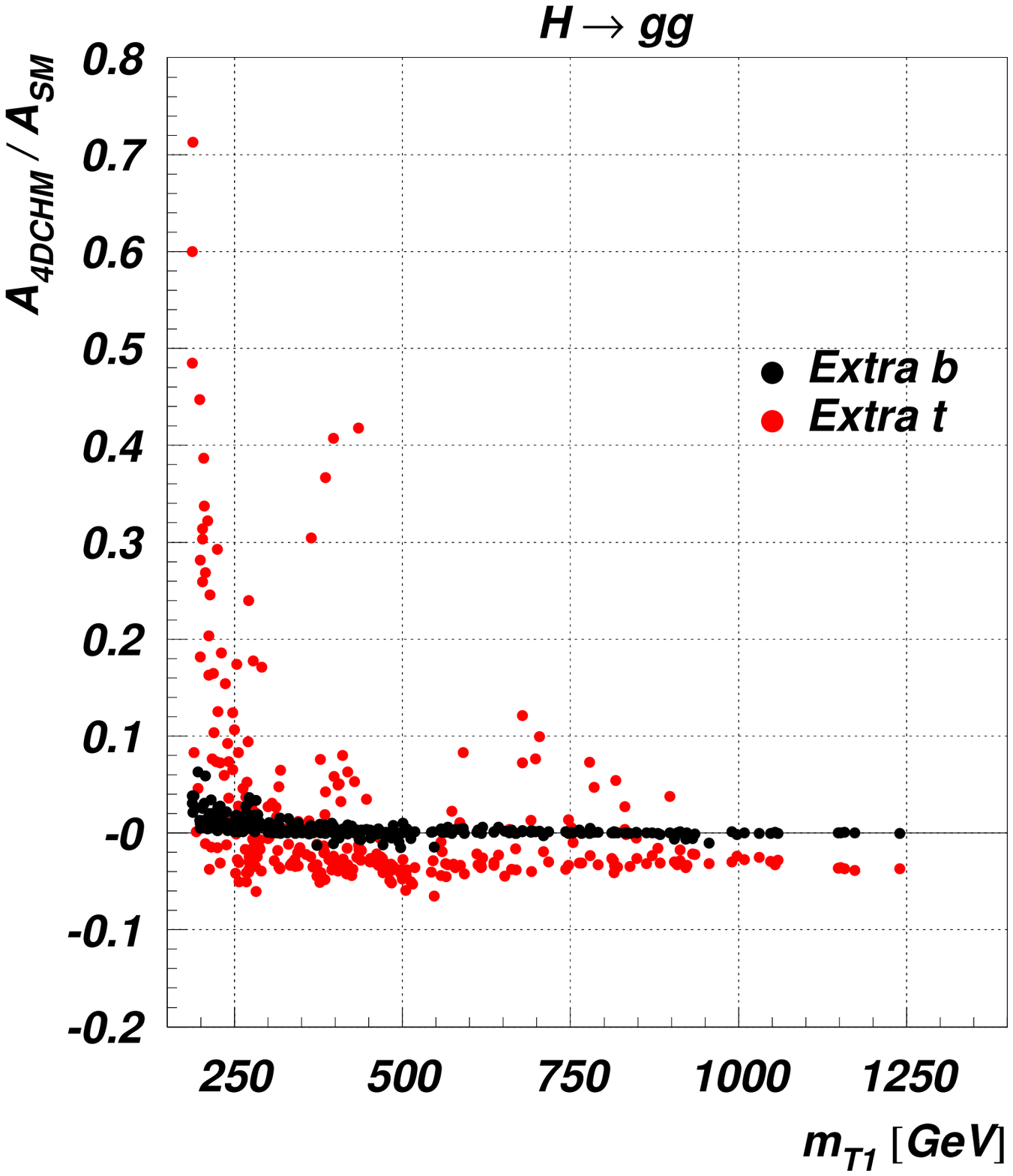}}
\subfloat[]{
\includegraphics[width=0.45\linewidth]{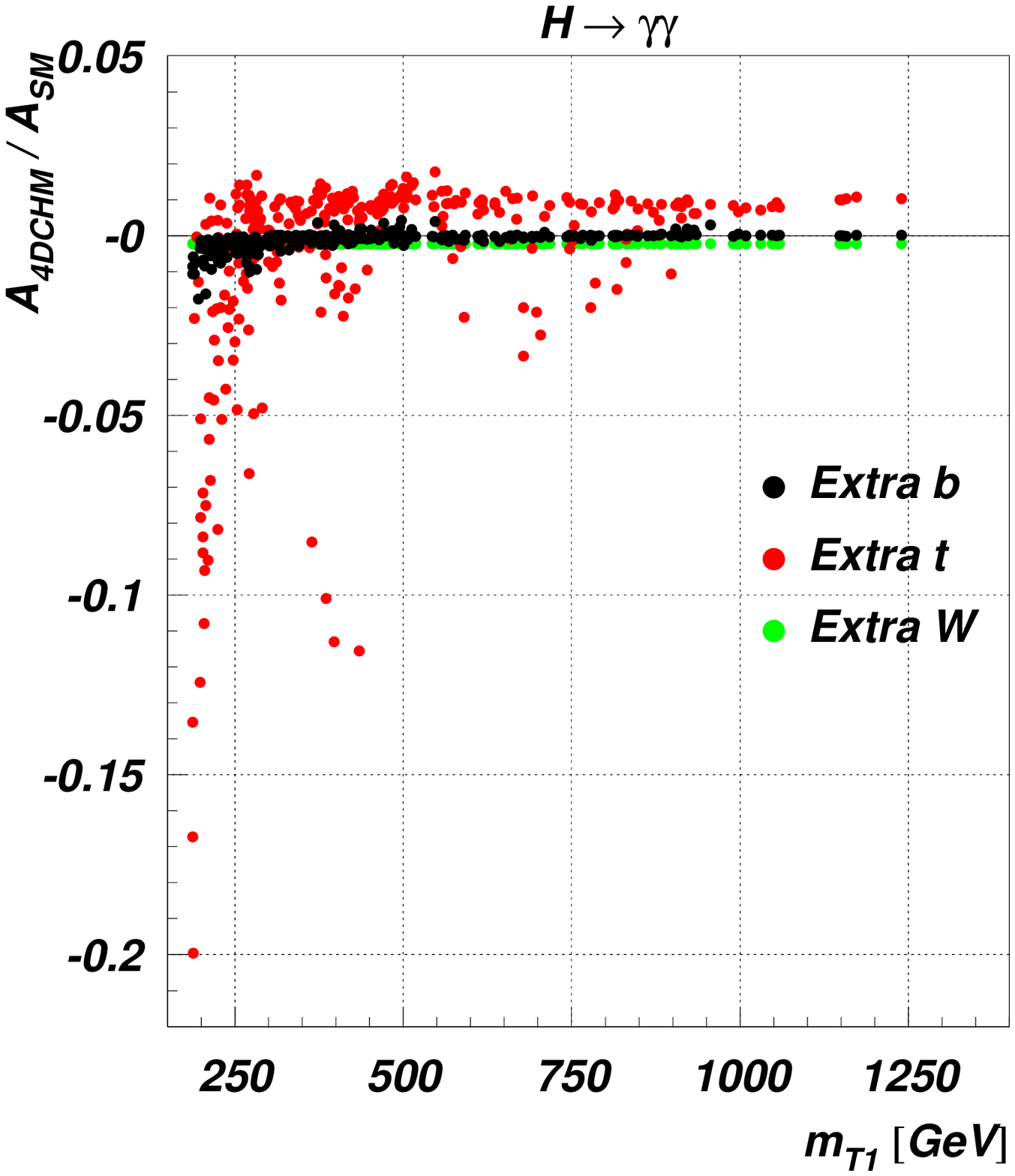}} \\
\subfloat[]{
\includegraphics[width=0.45\linewidth]{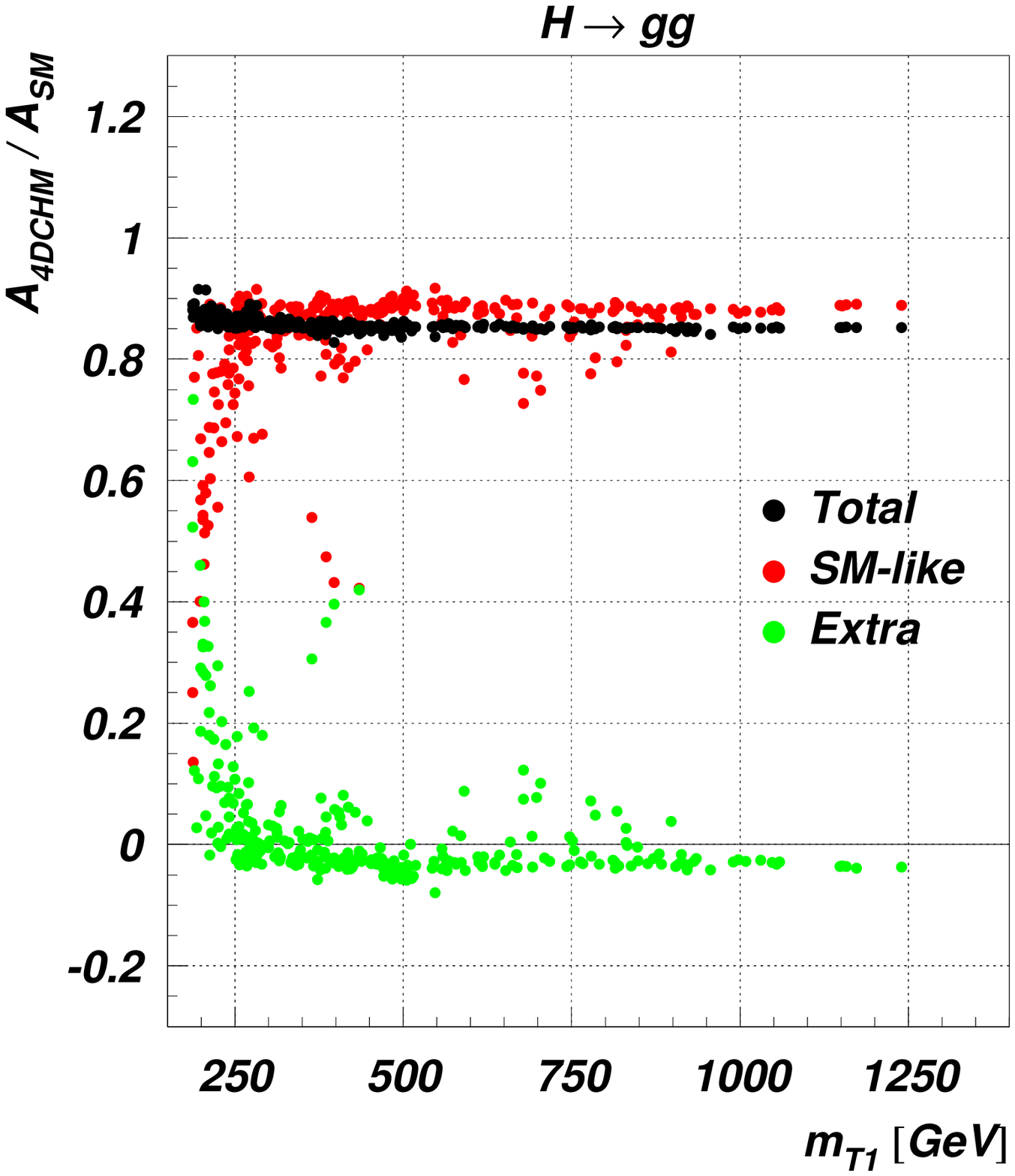}}
\subfloat[]{
\includegraphics[width=0.45\linewidth]{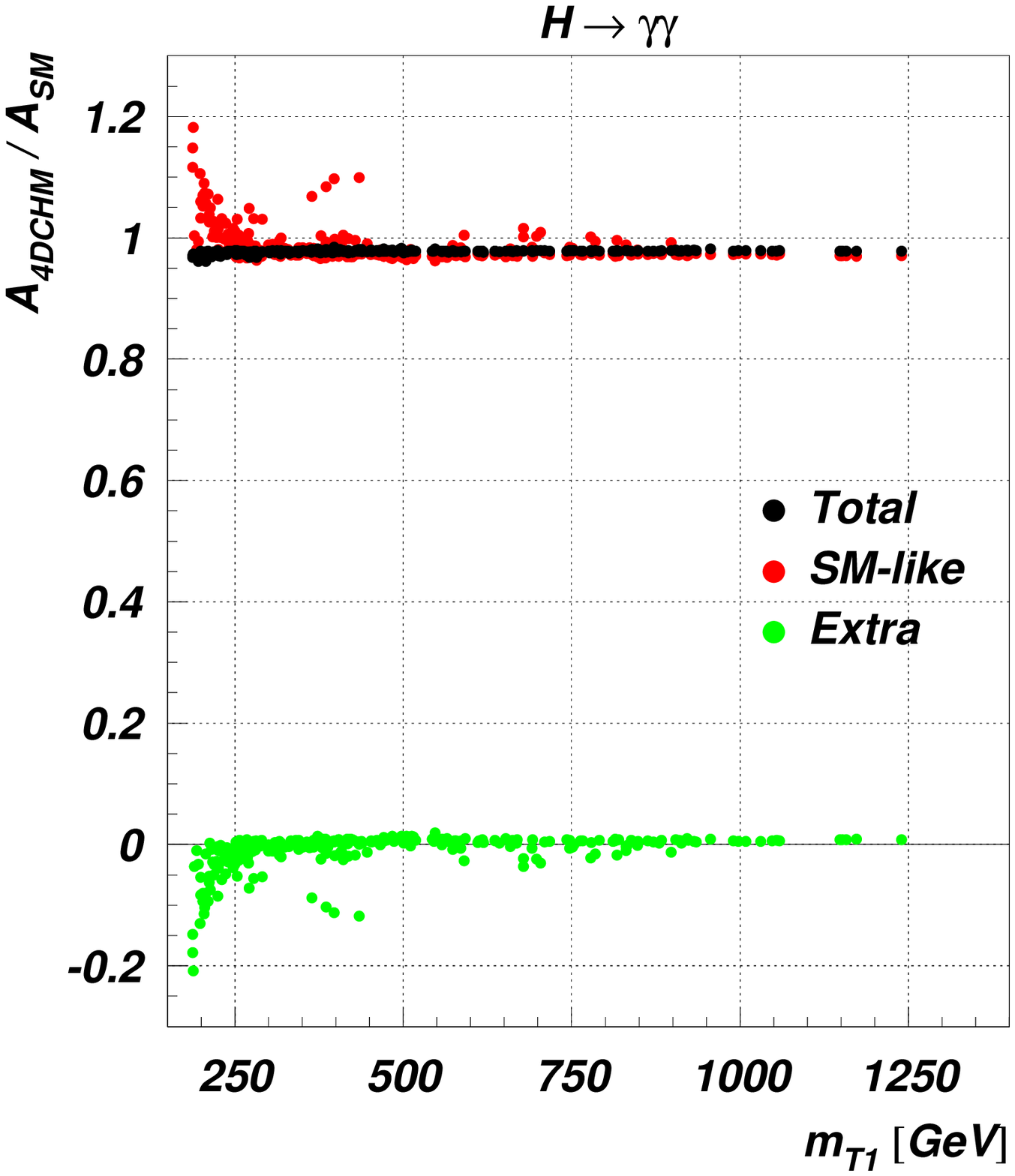}}
\caption{Loop-induced contributions to the amplitudes for $H\to gg$ (left panel) and $\gamma \gamma$ (right panel) in the 4DCHM normalised by the SM amplitude and plotted as a function of $m_{T_1}$. The top panels show only the genuine 4DCHM terms from the collective $W'$, $t'$ and $b'$ while the bottom panels show their sum against the SM-like terms due to $W$, $t$ and $b$ as well as the total. Here, $f=0.8$ TeV and $g_*=2.5$. \label{fig:amplitudes}}
\end{figure}

In Fig.~\ref{fig:amplitudes} we show the loop induced contributions to the amplitudes for the $H\to gg$ and $H\to \gamma \gamma$ decays in the 4DCHM normalised by the SM total amplitudes. In the top frame we show the 4DCHM contributions to the $H\to gg$ and $H\to \gamma \gamma$  amplitudes as separate channels, limited to the whole categories of $W'$, $t'$ and $b'$ states. From both plots we understand that the set{s} of such contributions which provide a sizable modification to the overall amplitude  {are} due to extra $t'$ {and extra $b'$}.  However, from the bottom frames in the same figure we can infer that the overall contribution to the amplitude, both in production and decay, of the genuine 4DCHM states is small but not negligible with respect to the SM-like ones, due to $W$, $t$ and $b$ states. {As for the sum over the $t$ and $t'$ contributions (i.e., $t$ plus $T_i$ with $i=1,\dots ,8$),} we notice a compensation effect between the SM{-like} and the genuine 4DCHM contributions (note the specular distributions of the points labelled `SM-like' and `Extra' in the plots, respectively){, which is due to the sum rule for the $t'$ effective result established in \cite{Azatov:2011qy}}. {As for the sum over the $b$ and $b'$ contributions (i.e., $b$ plus $B_i$ with $i=1,\dots ,8$), we notice that, for low values of $m_{T_1}$, such a compensation does not take place (in general there is no established sum rule for the $b'$ effective result, again, see \cite{Azatov:2011qy}), leading to a slight increase(decrease) of the total $H\to gg$($H\to \gamma \gamma$) amplitude.} Finally,  while the individual additional gauge boson contributions are always small, this is not the case for the additional fermion ones, as we shall see now.

In Fig.~\ref{fig:extra_contr} {we plot three frames: the top one shows the overall contributions to the $H\to gg$ amplitude from $t$, $t'$s, $b$ and $b'$s (i.e., `SM-like' plus `Extra' in both cases), while in the bottom frames}, channel by channel, we separate each of the important contributions and we show the ones which contain at least one point of the generated set above a $\sim 1$\% effect{, for $t'$s (left) and $b'$s (right)}.
\begin{figure}[!t]
\centering
\subfloat[]{
\includegraphics[width=0.45\linewidth]{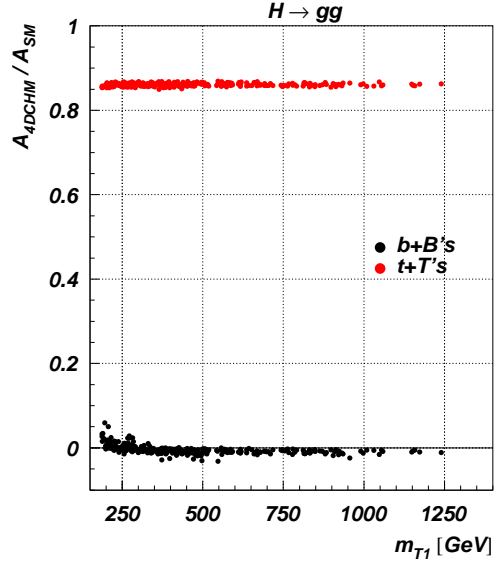}}\\
\subfloat[]{
\includegraphics[width=0.45\linewidth]{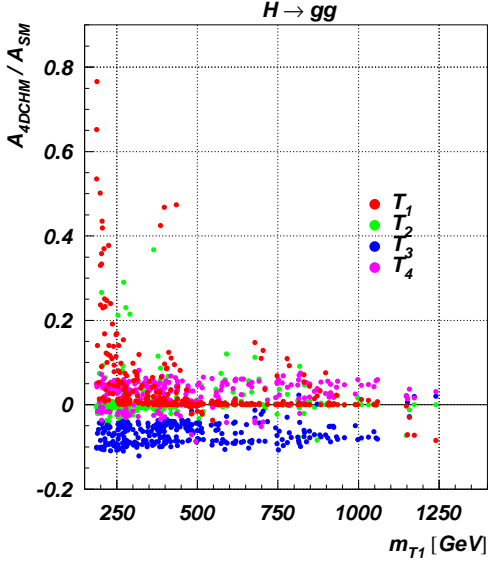}}
\subfloat[]{
\includegraphics[width=0.45\linewidth]{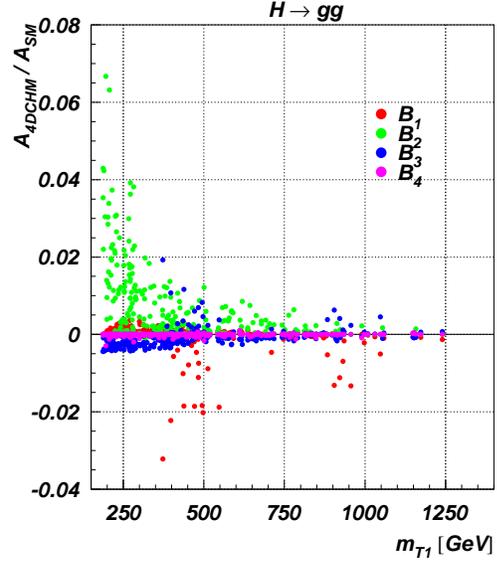}}
\caption{Loop-induced contributions to the amplitudes  for $H\to gg$ in the 4DCHM normalised by the SM amplitude and plotted as a function of $m_{T_1}$. Here the panels show the genuine 4DCHM terms which are individually relevant. Here, $f=0.8$ TeV and $g_*=2.5$. \label{fig:extra_contr}}
\end{figure}
Here,
by computation, we discover that only four among the $t'$s are able to produce substantial contributions: $T_1$, $T_2$, $T_3$ and $T_4$. From this figure we can associate specific effects to each one of the channels. First of all, we notice that $T_1$ and $T_2$ have always $0$ as a mean value while $T_3$ and $T_4$ have a {negative and positive} mean value, respectively, in the case of $H\to gg$. However, we underline the fact that the most sizable effect {from $t'$s,} for lower values of $m_{T_1}${,} is due to the contribution from $T_1$ itself, which is largely dominant. Furthermore, we highlight again that $T_3$ and $T_4$ produce two competing effects, but we also emphasise the fact that $T_3$ is always slightly dominant over $T_4$ and this tends to push the overall value against the SM one for the case of $H\to gg$. Further, as expected, the behaviour of the points plotted in the two bottom frames is (nearly) symmetric around $0$: this is due to the fact that they are weighted by the SM amplitude, for which the {largely} dominant contributions come from $t$ in the case of $H\to gg$ (whereas would have come from $W$ in the case of $H\to \gamma \gamma$). Finally, we notice that for points above the value $m_{T_1}\simeq 1.1$ TeV there is a clear inversion in the behaviour among the various channels, mainly due to threshold effects in the diagonalisation procedure of the fermionic mass matrices. 

Altogether then, individually large effects due to genuine 4DCHM contributions (specifically, of $t'$ type) entering with opposite signs at production and decay level cancel each other, so that their overall result is much smaller in comparison. 
{With regard to the $b'$ contributions, we notice that all of these have a mean value compatible with $0$. Among them, we see that the most important enhancement comes from the $B_2$ loop, which brings to deviations up to the $\sim 7$\%. Unlike the $t'$ case, we remark that the effective contributions from $b'$s do not respect any sum rule \cite{Azatov:2011qy}, so the enhancement from $B_2$ is not cancelled and it generates
substantial effects onto  the $H\to gg$ amplitude for low values of $m_{T_1}$.}

{In summary}, while the individual heavy fermion effects are at the ${\cal O}(1\%)$ level in the $H\to \gamma\gamma$ case, owing to the dominance of the $W$ term of SM origin, these are at the  ${\cal O}(10\%)$ in the $H\to gg$ case, where such a term is absent. While large effects onset by individual fermion contributions had been established in previous literature, see, e.g., Ref.~\cite{Azatov:2011qy}, it was not always appreciated therein that these would largely diminish in the presence of a complete and realistic mass spectrum, as the one generated here.

\clearpage
%\section*{Bibliography}
\bibliographystyle{h-physrev5}
\bibliography{biblio}

\end{document}